\documentclass[aps,twocolumn,superscriptaddress]{revtex4-1}
\pdfoutput=1

\usepackage{graphicx}
\usepackage{amsmath}
\usepackage{amssymb}
\usepackage{color}
\usepackage{bm}
\usepackage{subfigure}
\usepackage{appendix}

\definecolor{red}{rgb}{0.75,0,0}
\definecolor{blue}{rgb}{0,0,0.75}
\definecolor{green}{rgb}{0,0.5,0}

\DeclareMathOperator{\cn}{cn}
\DeclareMathOperator{\am}{am}

\begin{document}
\title{Polymorphism and bistability in adherent cells}

\author{Shiladitya Banerjee}
\affiliation{Department of Physics, Syracuse University, Syracuse, New York 13244, USA}

\author{Luca Giomi}
\affiliation{School of Engineering and Applied Sciences, Harvard University, Cambridge, Massachusetts 02138, USA}%
\affiliation{International School for Advanced Studies (SISSA), Via Bonomea 265, 34136 Trieste, Italy}

\begin{abstract}
The optimal shapes attained by contractile cells on adhesive substrates are determined by the interplay between intracellular forces and adhesion with the extracellular matrix. We model the cell as a contractile film bounded by an elastic cortex and connected to the substrate via elastic links. When the adhesion sites are continuously distributed, optimal cell shape is constrained by the adhesion geometry, with a spread area sensitively dependent on the substrate stiffness and contractile tension. For discrete adhesion sites, equilibrium cell shape is convex at weak contractility, while developing local concavities at intermediate values of contractility. Increasing contractility beyond a critical value, controlled by mechanical and geometrical properties of adhesion, cell boundary undergoes a discontinuous transition to a star-shaped configuration with cusps and protrusions, accompanied by a region of bistability and hysteresis.
\end{abstract}
\maketitle
\section{Introduction}
Mechanical force generation during cell-matrix adhesion is strongly influenced by the ability of cells to actively probe the mechanical and geometrical cues in the extracellular matrix~\cite{Discher2005}. Matrix stiffness plays a profound role in regulating a variety of cellular processes, from morphogenesis, motility to cell spreading and cytoskeletal activity. Cells adhering to softer substrates spread less and prefer to have well rounded morphologies, while they are more likely to exhibit branched patterns on stiffer substrates with greater spread area~\cite{Yeung2005,Chopra2011}. Experiments on micro-patterned adhesive islands revealed that cell fate, proliferation and spreading sensitively depend on adhesion geometry~\cite{Chen1997}. However, cellular response to extracellular determinants is strongly linked to myosin dependent activity of the cell cytoskeleton~\cite{Galbraith1998}. While myosin activity can influence force transmission by regulating the growth of focal adhesions~\cite{Riveline2001}, it can also drive changes in cell morphology, as seen by pharmacologically disrupting the cell cytoskeleton~\cite{Barziv1999,Asano2009} or by inhibiting myosin-II activity~\cite{Thery2006}. Traction forces exerted by cells on substrates can now be determined accurately using traction force microscopy or micropillar arrays~\cite{Pelham1997,duroure2005}, but the feedback between cell morphology and mechanics during adhesion to a matrix requires further theoretical investigation. In this article we present a minimal mechano-geometric model for isolated adherent cells that addresses a fundamental question in cell mechanics and morphogenesis: How intercellular and extracellular forces cooperate to control the geometry of cell shapes? At time scales when the cell is fully spread and develops stronger focal adhesions, the dominant forces in the cell stem from surface tension induced by actomyosin contractility and elasticity in the actomyosin cortex. These intracellular forces act in opposition to receptor-mediated adhesive forces in determining optimal cell shapes~\cite{Lecuit2007,Mader2007}. Although chemical pathways can trigger a feedback between cell activity and cell-substrate adhesion~\cite{Buchsbaum2007}, we instead focus on their mechanical cooperativity in regulating cell shapes. Tuning stiffness of the matrix and acto-myosin contractility, we discuss how cells can be driven through a series of morphological transitions - convex, concave, cusps and protrusions with associated hysteresis. In addition, we provide several analytical results relating geometrical properties of cells e.g. curvature, spread radius to mechanical properties such as substrate stiffness and contractile surface tension, that are amenable to experimental verification and quantitative comparison. 

The paper is organized as follows. In section 2, we introduce a minimal free energy model describing the optimal shape of the contact line of an adherent cell.  We retain three major contributions to the free energy stemming from : intracellular contractility, bending elasticity in the cell contour and adhesion to an extracellular substrate. We then proceed to study optimal solutions of the free energy for two distinct cases : (1) continuously distributed adhesion sites and (2) discrete adhesion sites. For continuously distributed adhesion sites, optimal shape of the cell contact line is always constrained by the geometry of the adhesion patch. We explicitly provide solution for the spread area of a cell constrained by a circular adhesion area, and analyze its dependence on substrate stiffness (Fig.~\ref{fig:spreading}). The result is in excellent qualitative and quantitative agreement with experimental trends. For discrete adhesion sites, cell contour develops non-uniform curvatures along its non-adherent segments. For low contractility and softer adhesions, the cell contour has a convex shape and one can analytically describe the shape and calculate traction forces transmitted to the substrate. In section 3 we discuss that by tuning contractility and substrate stiffness cell contour can be guided through a series of morphological transitions. Furthermore at intermediate values of  substrate stiffness (see Fig.~\ref{fig:phase}), increasing contractility beyond a critical value leads to formation of cusps and protrusions at adhesion sites. This transition involves a discontinuous change of the cell geometry characterized by a jump in the turning number of the contact line. The transition is further accompanied by a region of bistability and hysteresis in the dependence of cell perimeter on contractility. This result indicates how strongly intracellular and extracellular forces can control geometric properties of an adherent cell.

\section{Contractile Film Model for Adherent Cells}
We consider a thin film of an adherent cell subject to internal contractile forces. The shape of the cell contact line is parametrized by the contour ${\bm r}(s)$, where $s$ represents arc-length. The total mechanical energy of the cell can be approximated, on the basis of symmetry arguments, in the form:
\begin{equation}\label{eq:energy}
E = \sigma \int dA+ \oint ds\,\left(\alpha\kappa^{2}+\lambda\right)+k_{s}\oint ds\,\rho\,|\bm{r}-\bm{r}_0|^{2}\;,
\end{equation}
where $\sigma$ is the {\em effective} surface tension in the cell due to cytoskeletal contractility, $\kappa$ is the local curvature of the cell boundary, $\alpha$ the associated bending rigidity and $\lambda$ represents line tension at the cell boundary. The last term in Eq.~\eqref{eq:energy} represent the strain energy induced by the cell on a substrate of stiffness $k_s$ through focal adhesions localized at the cell edge~\cite{Wozniak2004} with density $\rho(s)$, so that the total number of adhesions is $N_{A}=\oint ds \rho$. For cells adhering to a thin continuous substrate (see Sec. \ref{sec:continuous}), $\bm{r}_{0}$ can be considered as the position of the cell boundary once the cell is fully spread and forces are predominately contractile, while for cells cultured on elastomeric pillars (see Sec. \ref{sec:discrete}), this is simply the pillar's rest position at the adhesion points. In the analytical framework presented here, we will treat the reference shape as an adjustable parameter to investigate different experimental situations.

The model assumes that the overall effect of acto-myosin contractility, that pulls the cell contour inwards reducing its contact area with the substrate, can be described by an {\em effective} surface tension $\sigma$. Thus the first term in Eq. \eqref{eq:energy} should not be interpreted as the classic hydrostatic tension that occurs at the interface between two fluids, but as an {\em active} normal stress resulting from the action of the motors. In order to estimate the order of magnitude of $\sigma$, we assume that the {\em active} myosin motors cross-linked with the cortical F-actin gel of mean thickness $h \simeq 0.1\ \mu$m, are distributed with an average areal density $\rho_m\simeq 10^4 \mu m^{-2}$, with effective stiffness $k_m\simeq 1$ pN/nm and mean stretch $\Delta_m \simeq 1$ nm~\cite{Howard2001}. Surface tension $\sigma$ can then be estimated as $\sigma\simeq h\rho_m k_m \Delta_m \simeq$ 1 nN/$\mu$m. This estimate comes to the same order of magnitude as reported for endothelial cells~\cite{Lemmon2005,Bischofs2009} and epithelial cells~\cite{Mertz2012}.

The second term in Eq. \eqref{eq:energy} describes the elasticity of the cell cortex. This consists of a bending energy density $\alpha\kappa^{2}$, reflecting the resistance of cortical actin in response of a change in curvature, and an effective line tension $\lambda$ that, similarly to the bulk tension $\sigma$, embodies the contractile forces due to the actin fibers lining the cell periphery~\cite{Bischofs2008,Bischofs2009}. The Euler-Lagrange equations for the shape that minimizes the energy \eqref{eq:energy} can be derived with standard methods~\cite{Mumford1993,Giomi2012}. This yields:
\begin{equation}\label{eq:general-euler-lagrange}
\alpha\left(2\kappa''+\kappa^{3}\right)-\lambda\kappa-\sigma+2k_{s}\rho(\bm{r}-\bm{r}_{0})\cdot\bm{n} = 0
\end{equation}
Here prime denotes derivative with respect to arc-length $s$ and $\bm{n}=\bm{r}''/|\bm{r}''|$ is the normal vector. Eq. \eqref{eq:general-euler-lagrange} expresses the balance between the total stress acting on a cross-section of the cortex and the body force $\bm{K}=2k_{s}\rho(\bm{r}-\bm{r}_{0})$ due to adhesion: 
\begin{equation}
\frac{d}{ds}(\bm{F}+\bm{\varSigma}+\bm{\varLambda})+\bm{K}=0
\end{equation}
where $\bm{F}=\alpha\kappa^{2}\bm{t}+2\alpha\kappa'\bm{n}$ (with $\bm{t}$ the tangent vector) is the elastic stress resultant, $\bm{\varSigma}=-\sigma[(\bm{r}\cdot\bm{t})\bm{n}-(\bm{r}\cdot\bm{n})\bm{t}]$ is the stress contribution of bulk contractility and $\bm{\varLambda}=-\lambda\bm{t}$ that of peripheral contractility.

Previous theoretical models~\cite{Barziv1999,Bischofs2008,Bischofs2009} have analyzed the competition of bulk and peripheral contractility and ignored the bending elasticity of the actin cortex (i.e. $\alpha=0$). In analogy with the Laplace law of capillarity, the steady state cell contour is then described by concave circular arcs of radius $\lambda/\sigma$ connecting adhesion sites. Here we focus on the opposite limit and consider the regime in which the force balance is dominated by the competition between cortex elasticity and bulk contractility, while the effect of peripheral contractility is negligible (i.e. $\lambda=0$). In this scenario, the curvature is generally non-uniform, especially in the neighborhood of adhesion sites. As we will see in the remainder of this article, incorporating bending elasticity leads to an extremely rich polymorphism and allows for a transition from purely convex to purely concave cell shape reminiscent of that observed in experiments on cardiac myocytes~\cite{Chopra2011}. Alternative models for cellular geometry and mechanics include a growing class finite element models \cite{Deshpande2006,Loosli2010,Farsad2012}, continuum mechanical models~\cite{Banerjee2011,Edwards2011} or network models~\cite{Torres2012}.

The Contractile Film Model defined by Eq. \eqref{eq:energy} is inspired by a classic problem in mechanics: finding the optimal shape of a capillary film bounded by an elastic rod. This problem was formulated by L\'evy \cite{Levy1884} in 1884 and for over a century it drew the attention of many researchers \cite{Tadjbakhsh1967,Flaherty1972,Arreaga2002,Vassilev2008,Djondjorov2011,Mora2012,Giomi2012} due to its tremendous richness of polymorphic and multi-stable behaviors. Unlike this simple system consisting of a film spanning an elastic boundary, however, the model proposed here for adhering cells does not involve any constraint on the length of the boundary, which is then only {\em softly constrained} by the adhesion with the substrate. This feature, introduces in the model a number of crucial mechanical properties, including an adaptive bending stiffness of the cell boundary.

\subsection{\label{sec:continuous}Continuous adhesions} 
\begin{figure}[t]
\centering
\includegraphics[width=0.8\columnwidth]{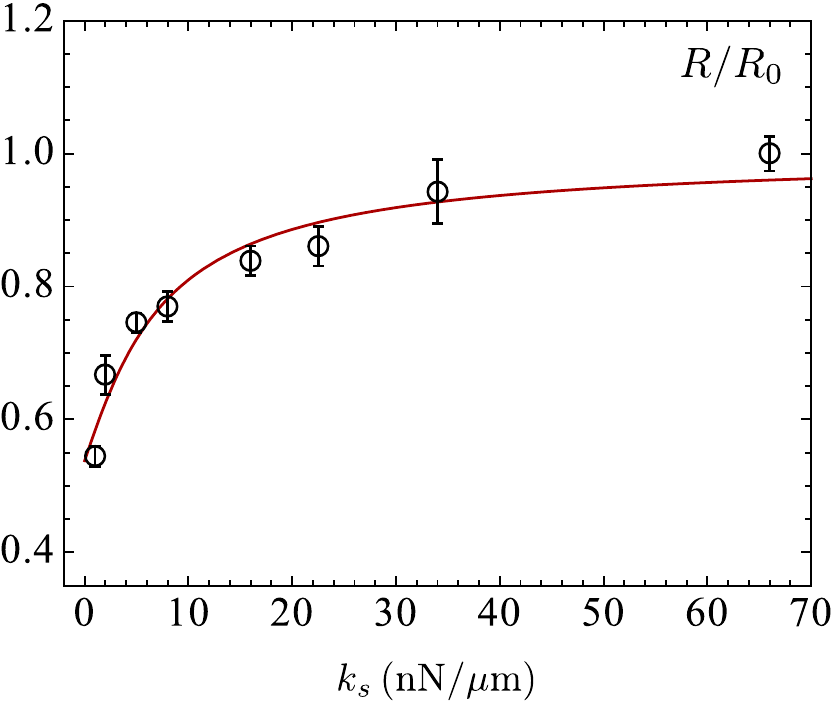}
\caption{\label{fig:spreading} Relative cell size $R/R_0$ as a function of substrate stiffness $k_s$ (solid black circles) for smooth muscle cells, 4 hours after plating on continuous elastic gels~\cite{Engler2004}. Cell radius is estimated from the projected cell area reported in~\cite{Engler2004} as $R=\sqrt{\text{area}}/\pi$. Substrate stiffness $k_s$ is determined from substrate Young's modulus $E_s$ as : $k_s=a E_s$, where $a$ is the characteristic focal adhesion size, with $a\sim 1\ \mu m$. Solid (red) line represents the solution to Eq.~\eqref{eq:radius} with with $\sigma= 1.05$ nN/$\mu$m and $\alpha/R_0^3=0.16$ nN/$\mu$m.}
\end{figure}
In this case the periphery of the cell forms contact with a single continuous adhesion site, so that $\rho=1/\mathcal{L}$ with $\mathcal{L}=\oint ds$ the perimeter of the cell. In presence of a uniform and isotropic substrate, we can assume the reference configuration to be a circle of radius $R_{0}$ so that a natural minimizer of the energy \eqref{eq:energy} would be a circle or radius $R$. Thus, setting $\lambda=0$, $\kappa=R^{-1}$ and $\rho^{-1}=2\pi R$ in Eq. \eqref{eq:general-euler-lagrange} yields the following cubic equation:
\begin{equation}\label{eq:radius}
(k_{s}+\pi\sigma)R^{3}-k_{s}R_{0}R^{2}-\pi\alpha = 0\;,
\end{equation}
The equation contains two length scales, $R_{0}$ and $\xi=(\alpha/\sigma)^{1/3}$, and a dimensionless control parameter $k_{s}/\sigma$ expressing the relative amount of adhesion and contraction. For very soft anchoring $k_{s}\ll \sigma$ and Eq.~\eqref{eq:radius} admits the solution $R=\xi$. Thus in non-adherent cell segments, corresponding to the limit $k_s=0$, radius of curvature scales with surface tension as $R \sim \sigma^{-1/3}$. The same scaling law is also predicted using \textit{active cable network} models of an adherent cell~\cite{Torres2012}. If the cell is rigidly pinned at adhesion sites, $k_{s}\gg \sigma$ and $R\rightarrow R_0$. For intermediate values of $k_{s}/\sigma$ the optimal radius $R$ interpolates between $\xi$ and $R_{0}$ and is an increasing function of the substrate stiffness $k_{s}$, in case $\xi<R_{0}$, or a decreasing function if $\xi>R_{0}$. For $\xi=R_{0}$, the lower and upper bound coincide, and the solution is $R=R_0$. In particular, the case $R_0>\xi$ reproduces the experimentally observed trend that cell projected area increases with increasing substrate stiffness before reaching a plateau at higher stiffnesses~\cite{Engler2004,Yeung2005,Chopra2011}. We fit the solution to Eq.~\eqref{eq:radius} to the measured projected areas of smooth muscle cells (SMCs) adhering to continuous elastic gels of varying substrate elastic modulus~\cite{Engler2004}, as shown in Fig.~\ref{fig:spreading}. Data for the spread area of SMCs are taken 4 hours after plating onto the substrate, when they retain rounded morphologies. The fitted value for surface tension $\sigma=1.05$ nN/$\mu$m comes to the same order of magnitude as reported for endothelial cells~\cite{Lemmon2005,Bischofs2009}, epithelial cells~\cite{Mertz2012} and is consistent with the numerical estimate provided earlier. The fit also provides a value for the bending rigidity $\alpha=4.62 \times 10^{-16}$ Nm$^2$. The asymptotic behavior and various limits of the solution are well captured by the interpolation formula:
\begin{equation}
R\approx\frac{k_{s}R_{0}+3\pi\sigma\,\xi}{k_{s}+3\pi\sigma}
\end{equation}
indicating that larger surface tension, hence larger cell contractility $\sigma$ leads to lesser spread area, consistent with the experimental observation that myosin-II activity retards the spreading of cells~\cite{Wakatsuki2003}. Standard stability analysis of this solution under a small periodic perturbation in the cell radius shows that the circular shape is always stable for any values of the parameters $\sigma$, $k_{s}$ and $R_{0}$.
\subsection{\label{sec:discrete}Discrete adhesions}
\begin{figure}[t]
\centering
\includegraphics[width=\columnwidth]{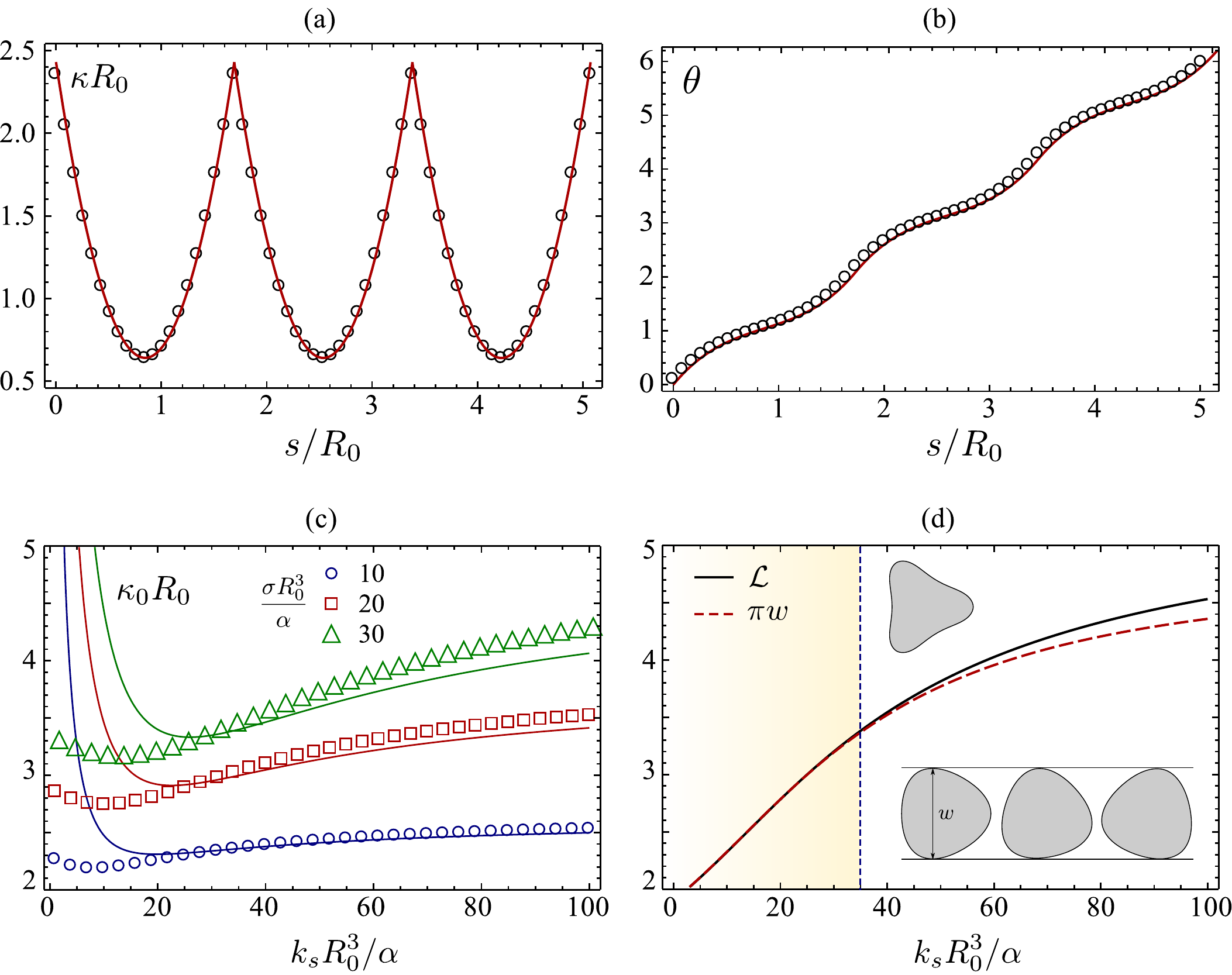}
\caption{\label{fig:geometry}Cell anchored onto three pointwise adhesions located at the vertices of an equilateral triangle. The curvature (a) and the tangent angle (b) as function of arc-length for $\sigma R_{0}^{3}/\alpha=10$, $k_{s}R_{0}^{3}=50$ and $N_{A}=3$. The circles are obtained from a numerical minimization of a discrete version of the energy \eqref{eq:energy}, while the solid lines corresponds to our analytical approximation. (c) The total cell length $\mathcal{L}$ as a function of adhesion stiffness. For small stiffnesses the cell boundary form a curve of constant width (lower inset) and $\mathcal{L}=\pi w$, with $w$ the width of the curve. This property breaks down for larger stiffnesses when inflection points develops (upper inset). (d) The curvature $\kappa_{0}$ at the adhesion points as a function of the substrate stiffness for various contractility values. The points are obtained from numerical simulations while the solid lines correspond to our analytical approximation.}
\end{figure}
\begin{figure*}
\centering
\includegraphics[width=0.9\textwidth]{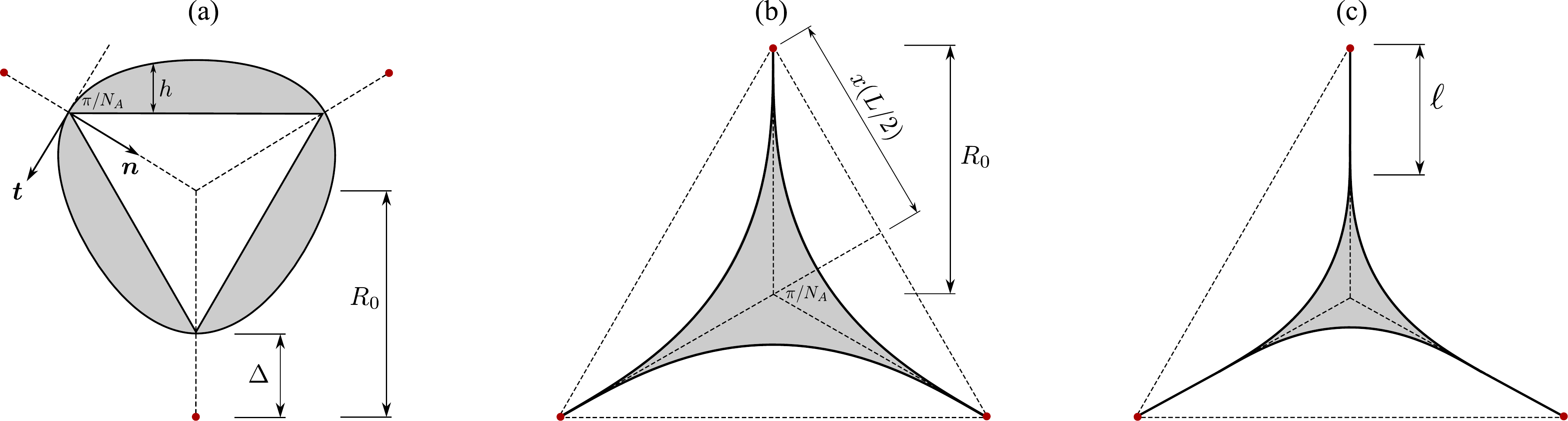}
\caption{\label{fig:schematic}Cell anchored onto three pointwise adhesions located at the vertices of an equilateral triangle. (a) $\sigma < \sigma_{c1}$, cell contour is everywhere convex with constant width. (b) $\sigma=\sigma_p$, cell contour is purely concave with cusps at adhesion points and without protrusions. (c) $\sigma>\sigma_{c2}$, cusps are connected to the substrate by means of a protrusion of length $\ell$.}
\end{figure*}
For cells adhering to discrete number of adhesion sites, one can show that the circular solution for the cell boundary is never stable and there is always a non-circular configuration with lower energy. For simplicity, we assume that $N_A$ adhesion sites are located at the vertices of a regular polygon of circumradius $R_{0}$, with density $\rho(s)=\sum_{i=0}^{N_{A}-1}\delta(s-iL)$,  and $L$ the distance between subsequent adhesions.  Optimal cell shape is given by the solution of the equation:
\begin{equation}\label{eq:euler-lagrange}
\alpha\left(2\kappa''+\kappa^{3}\right)- \sigma + 2 k_s\sum_{i=0}^{N_{A}-1}  \delta(s-iL)\left({\bm r}-{\bm r}_0\right)\cdot {\bm n}= 0\;.
\end{equation}
Due to the $N_{A}$-fold symmetry of the adhesion sites, adhesion springs stretch by an equal amount $\Delta$ in the direction of the normal vector: $(\bm{r}_{i}-\bm{r}_{0i})\cdot\bm{n}_{i}=\Delta$, $i=1,\,2\ldots N_{A}$. As a consequence of the localized adhesion forces, the curvature is non-analytical at the adhesion points. Integrating Eq. \eqref{eq:euler-lagrange} along an infinitesimal neighborhood of a generic adhesion point $i$, one finds the following condition for the derivative of the curvature at the adhesion points:
\begin{equation}\label{eq:discontinuity}
\kappa'_{i}=-\frac{k_{s}}{2\alpha}\,\Delta\;.
\end{equation}

The local curvature of the segment lying between adhesion points is on the other hand determined by the equation $\alpha\left(2\kappa''+\kappa^{3}\right)-\sigma=0$, with the boundary conditions : $\kappa(iL)=\kappa((i+1)L)=\kappa_0$. Without loss of generality we consider a segment located in $s\in[0,L]$. Although an exact analytic solution this nonlinear equation is available (see Ref. \cite{Vassilev2008} and Appendix C), an excellent approximation can be obtained by neglecting the cubic nonlinearity (Fig.~\ref{fig:geometry}a-b). With this simplification, Eq. \eqref{eq:euler-lagrange} admits a simple solution of the form:
\begin{equation}\label{eq:curvature}
\kappa(s)=\kappa_0 + \frac{\sigma}{4\alpha}\,s(s-L)\;.
\end{equation}
Eqs. \eqref{eq:curvature} and \eqref{eq:discontinuity} immediately allow us to derive a condition on the cell perimeter: $L=2k_{s}\Delta/\sigma$. Furthermore, the latter condition leads to a linear relation between traction force $T= 2 k_s \Delta$, and cell size :
\begin{equation}
T=\sigma L\;,
\end{equation}
which is indeed observed in traction force measurements on large epithelial cells~\cite{Mertz2012}.

To determine the end-point curvature $\kappa_{0}$, we use the turning tangents theorem for a simple closed curve \cite{Gray1997}, which requires $\int_{0}^{L}ds\,\kappa = 2\pi/N_{A}$. This leads to following relation between local curvature and segment length, or equivalently traction force, at the adhesion sites :
\begin{equation}
\kappa_{0}
=\frac{\sigma L^{2}}{24\alpha} + \frac{2\pi}{N_{A}L}
=\frac{T^{2}}{24\,\alpha\sigma} + \frac{2\pi\sigma}{N_A T}\;.
\end{equation}
A plot of $\kappa_{0}$ as a function of the substrate stiffness is shown in Fig. \ref{fig:geometry}c.

Finally, to determine the optimal length of the cell segment $L$, we are going to make use of a remarkable geometrical property of the curve obtained from the solution of Eq.~\eqref{eq:euler-lagrange} with discrete adhesions: the fact of being a {\em curve of constant width} \cite{Gray1997}. The width of a curve is the distance between the uppermost and lowermost points on the curve (see lower inset of Fig. \ref{fig:geometry}d). In general, such a distance depends on how the curve is oriented. There is however a special class of curves, where the width is the same regardless of their orientation. The simplest example of a curve of constant width is clearly a circle, in which case the width coincides with the diameter. A fundamental property of curves of constant width is given by the Barbier's theorem \cite{Gray1997}, which asserts that the perimeter $\mathcal{L}$ of any curve of constant width is equal to width $w$ multiplied by $\pi$: $\mathcal{L}=\pi w$. As illustrated in Fig.~\ref{fig:geometry}d, this is confirmed by numerical simulations for low to intermediate values for contractility and stiffness. With our setting, the cell width is given by:
\begin{equation}\label{eq:width}
w = (R_{0}-\Delta)(1+\cos\pi/N_{A})+h(L/2)\;,
\end{equation}
where $h(s)=\int_{0}^{s}ds'\,\sin\theta(s')$ is the height of the curve above a straight line connecting subsequent adhesions and
\begin{equation}
\theta(s)=\int_{0}^{s}ds'\,\kappa(s')=\theta_{0}+\kappa_{0}s+\frac{\sigma}{24\alpha}\,s^{2}(2s-3L)
\end{equation}
the angle formed by the tangent vector with the $x-$axis of a suitable oriented Cartesian frame (Fig.~\ref{fig:schematic}a). For small angles $h$ can be approximated as : $h(s) \approx s(L-s)\left[\pi/(N_{A}L)-(\sigma/48\alpha)\,s(L-s)\right]$. Using this together with Eq. \eqref{eq:width} and the Barbier's theorem with $\mathcal{L}=N_{A}L$ allow us to obtain a quartic equation for the cell length and the traction force, whose approximate solution is given by:
\begin{equation}\label{eq:traction}
T \simeq \frac{\sigma R_{0}}
{\left(g_0 + \frac{\sigma}{2k_s}\right)\left[1+\frac{7\sigma R_0^3}{\alpha g_1 } \left(g_0 + \frac{\sigma}{2k_s}\right)^{-4}\right]^{1/7}}\;,
\end{equation}
where, $g_0=(4N_A^2-\pi^2)/\left[4\pi N_{A} (1+\cos\pi/N_{A})\right]$ and $g_1=768(1+\cos\pi/N_{A})$. Eq.~\eqref{eq:traction} supports the experimental trend that traction force increases monotonically with substrate stiffness $k_{s}$ before plateauing to a finite value for higher stiffnesses~\cite{Ghibaudo2008,Mitrossilis2009}. The plateau value increases with increasing contractility (Fig.~\ref{fig:traction}a). Traction force grows linearly with increasing contractility for $\sigma R_0^3/\alpha \ll 1$, before saturating to the value $2k_s R_0$ at large contractility $\sigma R_0^3/ \alpha \gg 1$, as shown in Fig.~\ref{fig:traction}b. Eq.~\eqref{eq:traction} is also consistent with experimentally observed trend that reducing contractility by increasing the dosage of myosin inhibitor Blebbistatin, leads to monotonic drop in traction forces~\cite{Mitrossilis2009}.

In the calculation presented in this section we have neglected the contribution of peripheral contractility embodied in the effective line tension $\lambda$. From the point of view of force balance, increasing $\lambda$ has the effect of rotating the stress resultant toward the tangential direction. This creates a boundary layer between the adhesion points, where the curvature $\kappa_{0}$ is dictated by the balance between adhesion and bending, and the central region, where the curvature $\kappa\approx\sigma/\lambda$ is dominated by the balance between normal and tangential contractility. The size of the boundary layer is approximatively $\sqrt{\alpha/\lambda}$.

\section{Inflections, cusps and protrusions}
\begin{figure}[t]
\centering
\includegraphics[width=\columnwidth]{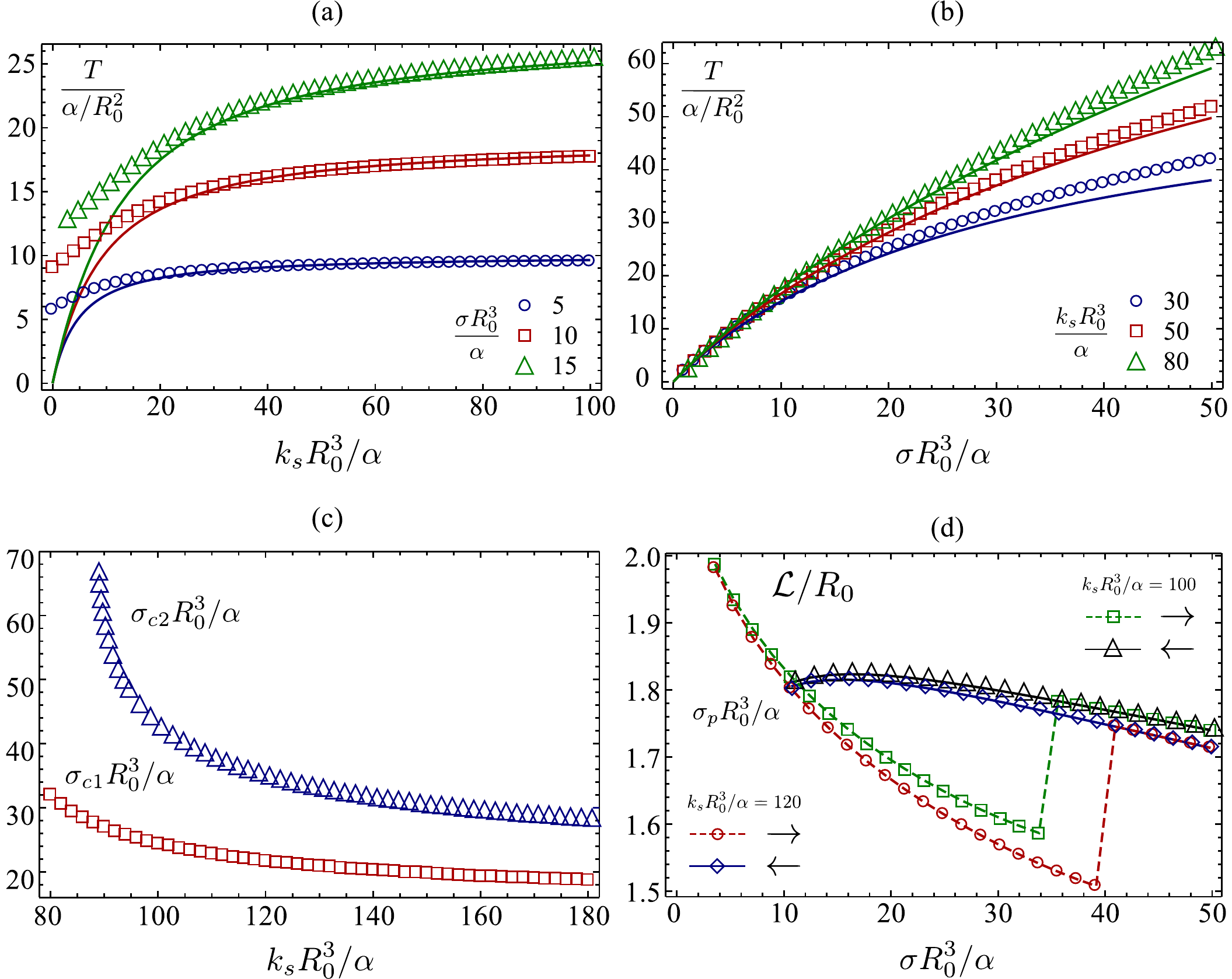}
\caption{\label{fig:traction}Traction force as a function of substrate stiffness (a) and contractility (b) obtained from a numerical minimization of a discrete analog of Eq. \eqref{eq:energy}. Solid curves denote the approximate traction values obtained from Eq. \eqref{eq:traction}. (c) Boundary length $\mathcal{L}$ obtained by increasing (squares) and then decreasing (triangles) the contractility for substrate stiffnesses $k_{s}R_{0}^{3}/\alpha=100$ (green squares, black triangles) and $k_{s}R_{0}^{3}/\alpha=120$ (red squares, blue triangles). The diagram shows bistability in the range $\sigma_p < \sigma < \sigma_{c2}$. (d) The critical contractility $\sigma_{c1}$ and $\sigma_{c2}$ as functions of substrate stiffness.}
\end{figure}

\begin{figure}[t]
\centering
\includegraphics[width=0.8\columnwidth]{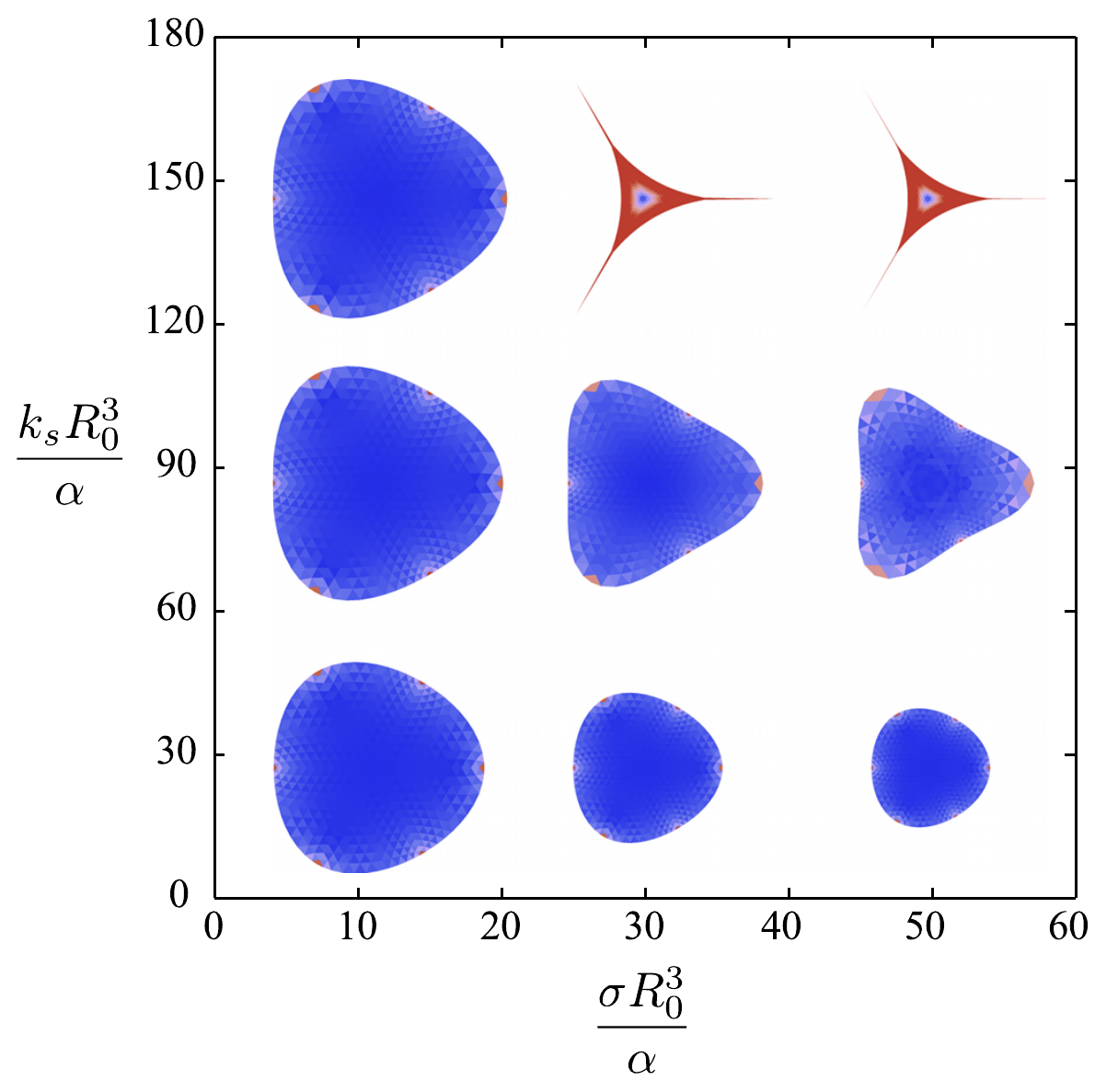}
\caption{\label{fig:phase}Phase diagram in $\sigma$-$k_s$ plane showing optimal configuration obtained by numerical minimization of the energy \eqref{eq:energy} for $N_A=3$.}
\end{figure}

For low to intermediate values of $\sigma$ and $k_s$, cell shape is convex and has constant width. Upon increasing $\sigma$ above a $k_{s}-$dependent threshold $\sigma_{c1}$, however, the cell boundary becomes inflected (see Fig. \ref{fig:phase} and upper inset of Fig. \ref{fig:geometry}d). Initially a region of negative curvature develops in proximity of the mid point between two adhesions, but as the surface tension is further increased, the size of this region grows until positive curvature is preserved only in a small neighborhood of the adhesion points. Due to the presence of local concavities, the cell boundary is no longer a curve of constant width. Convex and concave regions are separated by inflection points, given by the solution to $\kappa=0$, or explicitly: $s^{2}-Ls+4\alpha\kappa_{0}/\sigma=0$. In order for this equation to have real solutions one needs $\sigma L^3>96\pi \alpha/N_A$. Fig.~\ref{fig:traction}c shows $\sigma_{c1}$ as a function of $k_{s}$. Prior experimental studies~\cite{Thery2006,James2008} have indicated that lamellipodia formation is predominant along convex edges or sharp corners, whereas contractile stress fibers assemble along concave regions. Lamellipodia formation implies greater motile activity along those corners. On softer substrates, with weak adhesion, cell is more motile on average than on stiff substrates. This is because increasing substrate stiffness promotes formation of concave arcs along non-adherent sites, thus reducing the total area spanned by convex regions. As such, lamellipodia distribution is controlled by the geometry of adhesion sites for both continuous and discrete cases.
Upon increasing $\sigma$ above a further threshold value $\sigma_{c2}$, the inflected shape collapses giving rise to the star-shaped configurations shown in upper right corner of Fig.~\ref{fig:phase}. These purely concave configurations are made by arcs whose ends meet in a cusp. The cusp is then connected to the substrate by a protrusion consisting of a straight segment of length $\ell$ that extends until the adhesion point rest position, so that $\Delta\approx 0$ (Fig. ~\ref{fig:schematic}c) (see Appendix B). The cell boundary becomes pinned at adhesion sites as a result of having to satisfy force-balance, Eq.~\eqref{eq:euler-lagrange}, and adhesion-induced boundary condition, Eq.~\eqref{eq:discontinuity}, while accommodating large contractile tensions at its neighbourhood. This results in spontaneous expansion in the cell perimeter. Unlike the previous transition from convex to non-convex shapes, this second transition occurs discontinuously and is accompanied by a region of bistability in the range $\sigma_{p}< \sigma < \sigma_{c2}$, where $\sigma_{p}$ is the value of $\sigma$ at which the protrusions have zero length and the shape of the cell is that sketched in Fig. \ref{fig:schematic}b. This is clearly visible in the hysteresis diagram in Fig.~\ref{fig:traction}d showing the optimal length obtained by numerically minimizing a discrete analog of Eq. \eqref{eq:energy} in a cycle and using as initial configuration the output of the previous minimization. The onset of bistability is regulated by substrate stiffness as shown in Fig.~\ref{fig:traction}c, with stiffer substrates promoting transition to cusps at lower $\sigma_{c2}$. Away from the protrusion, the curvature has still the form given in Eq. \eqref{eq:curvature}, with $\kappa_{0}=0$ so that the boundary is everywhere concave or flat and the bending moment $\bm{M} = 2\alpha\kappa\bm{\hat{z}}$ does not experience any unphysical discontinuity at the protrusion's origin.

From the shape of the cell at $\sigma=\sigma_{p}$ we can construct all the shapes at $\sigma>\sigma_{p}$ by mean of a {\em similarity} transformation. To see this let us set $\ell=0$ at $\sigma=\sigma_{p}$ so that the shape of the cell will be of the kind illustrated in Fig.~\ref{fig:schematic}b. In the following we will refer to this as the {\em reference shape}. The approximated expression for the curvature is the same given in Eq. \eqref{eq:curvature}, but with $\kappa_{0}=\Delta=0$ and $\kappa'$ unconstrained since the last term in Eq. \eqref{eq:euler-lagrange} vanishes identically. The quantities $\sigma_{p}$ and the length $L_{p}$ of the reference shape are left to determine. To achieve this, a first condition can be obtained by observing that: $x(L_{p}/2)=R_{0}\sin\pi/N_{A}$, where $x(s)$ is the projection of the curve on the edge of the circumscribed polygon (see Fig.~\ref{fig:schematic}b). A second condition is given by the theorem of turning tangents for a simple closed curve with $N_{A}$ cusps: $\int_{0}^{L_{p}}ds\,\kappa=\pi(2-N_{A})/N_{A}$ (see Appendix B). In the case $N_{A}=3$, for instance, the right-hand side is equal to $-\pi/3$, corresponding to the fact that the tangent vector rotates clockwise by $60^{\circ}$ as we move counterclockwise along the curve from one cusp to the next. These allow us to approximate:
\begin{subequations}
\begin{gather}	
L_{p} \approx \frac{2N_{A}R_{0}}{\pi(N_{A}-2)}\,\sin\frac{\pi}{N_{A}}\;,\\[7pt] 	
\sigma_p\approx \frac{3\alpha\pi^4}{R_0^3 \sin^3\frac{\pi}{N_A}}\left(\frac{N_A-2}{N_A}\right)^4\;,
\end{gather}
\end{subequations}
which define the {\em reference shape} shown in Fig.~\ref{fig:schematic}b.

Next, following Ref. \cite{Flaherty1972,Djondjorov2011}, we notice that the force balance equation $2\kappa''+\kappa^{3}-\sigma/\alpha=0$ is invariant under the scaling transformation:
\begin{equation}
(s,\,\kappa,\,\sigma) \rightarrow \left(\Lambda\,s,\,\frac{\kappa}{\Lambda},\,\frac{\sigma}{\Lambda^{3}}\right)\;.
\end{equation}
Consequently, the equilibrium shape obtained for a given value of $\sigma>\sigma_{p}$ are similar to the reference shape with a scaling factor $\Lambda=(\sigma_{p}/\sigma)^{1/3}<1$. Accordingly, the closed curve is rescaled so that $L=\Lambda L_{p}$ and $A=\Lambda^{2}A_{p}$ with $A_{p}$ the area of the reference shape. This beautiful geometric property immediately translates into the following algorithm to construct shapes with protrusion (Fig. \ref{fig:schematic}c): {\em 1)} Given the surface tension $\sigma>\sigma_{p}$ we calculate the scaling factor $\Lambda$. {\em 2)} We rescale the reference curve so that $L=\Lambda L_{p}$. {\em 3)} Finally, we fill the distance between the adhesion points and the cusps with straight segments of length $\ell=R_{0}(1-\Lambda)$ (since $R_{0}$ is the circumradius of the reference shape and $\Lambda\,R_{0}$ that of the rescaled shape). This latter step, ultimately allows us to formulate a scaling law for the length of protrusions that can be tested in experiments:
\begin{equation}\label{eq:protrusion}
\ell/R_{0} = 1-(\sigma_{p}/\sigma)^{1/3}\;.	
\end{equation}

It should be stressed that our knowledge of the convex/concave transition is still very preliminary. This instability is different from the classical Euler buckling \cite{Love1927}, which originates from the trade-off between compression and bending and is a supercritical pitchfork bifurcation. The appearance of cusps is reminiscent, to some extent, of the {\em sulcification} instability in neo-Hookean solids \cite{Biot1965,Hohlfeld2011,Hohlfeld2012,Tallinen2012}, but there is far from being a precise mapping. One of the fundamental aspect that distinguishes our model form classical elasticity relies on the fact that the perimeter is not hardly constrained, but only subject to a soft constraint by mean of the adhesion springs. The length of an elastic object affects its overall flexibility (i.e. long filaments are floppy and easy to bend, while short filaments are stiff), thus, when the effective surface tension is increased, the whole cell boundary becomes shorter and stiffer. Because stiff materials are difficult to bend, but easy to break, a possible interpretation could be the following. For sufficiently large adhesion, increasing the surface tension has the effect of bending and stiffening the cell boundary in proximity of the adhesion sites, until, above a certain surface tension, the cell boundary is too stiff to continue bending and fractures. The cracks are localized at the adhesion points, where the curvature initially focuses, giving rise to the cusps observed in the simulations. However, a thorough understanding of this phenomenon remains a challenge for the future.

\section{Discussion}
The Contractile Film Model describes equilibrium cell shapes and does not account for the dynamics associated with adhesion remodeling and actin filament turnovers. Adhesion sites are static, with controllable density and spatial distribution, as can be best realized using micropatterning techniques~\cite{Thery2006}. The model provides a quantitative framework to describe how polymorphic cell shapes arise by tuning substrate stiffness, adhesion geometry and cell activity. The presence of bending deformations in the cell periphery naturally allows for optimal cell shapes with non-uniform boundary curvatures, a feature not included in previous theoretical works on cell shapes~\cite{Barziv1999,Bischofs2008,Bischofs2009}. Bending in the cell boundary can also arise due to splay deformations in the Arp2/3 regulated actin array in the lamellipodium. Although our model relies on local mechanotransduction through adhesions localized at the cell edge, in reality traction stresses penetrate inside the cell up to a characteristic depth controlled by cellular and substrate stiffness~\cite{Mertz2012}. Our model is thus applicable to cell sizes much larger than traction penetration depth and predicts the same trend on the dependence of traction forces on substrate stiffness as derived using long-range elastic models~\cite{Banerjee2012}. Although, local mechanosensing at cell periphery coupled with global surface tension due to cytoskeletal contractility can accurately capture experimental trends for cell size and traction forces, the effect of non-local interactions of the cytoskeleton with the substrate cannot be neglected at actin remodeling time scales.

An important consequence of increasing surface tension is the loss of stability of smooth shapes and a discontinuous transition to cusps and protrusion (Fig. \ref{fig:traction} and \ref{fig:phase}). The transition is favored on stiffer substrates (see Fig. \ref{fig:phase}) and leads to spontaneous expansion in the cell perimeter and relaxation of localized adhesion springs. Such a transition could also possibly occur on cellular timescales via chemo-mechanical instabilities induced by coupling of motor activity with ligand-receptor kinetics at adhesion sites. Instead here it emerges as a consequence of the cell boundary satisfying of energy minimization and global geometrical constraint imposed by the theorem of turning tangents. To our knowledge no experimental evidence has yet been put forward of such instabilities. The result can be tested in cell traction assays by varying motor activity in a cycle.

Finally, recent experiments on multicellular systems~\cite{Mertz2012} demonstrated that cohesive cell colonies behave like single cells in their distribution of traction stresses and presence of supracellular actin cables localized to the colony periphery. The Contractile Film Model can be conveniently used to study shapes of strongly coupled cell colonies, where colony surface tension stems from actomyosin contractility as well as strength of cadherins mediating cell-cell adhesions.

\begin{acknowledgments}
We thank Cristina Marchetti for many useful discussions. SB acknowledges support from National Science Foundation through awards DMR-0806511 and  DMR-1004789. LG is supported by the NSF Harvard MRSEC, the Harvard Kavli Institute for Nanobio Science \& Technology and the Wyss Institute.
\end{acknowledgments}

\section*{Appendix A $-$ Numerical Simulations}
The data shown in Figs. \ref{fig:spreading} and \ref{fig:traction}a,b,d have been obtained by numerically minimizing the following discrete version of the energy \eqref{eq:energy}:
\begin{equation}\label{eq:discrete-energy}
E_{1} = \frac{\sigma}{2}\sum_{i=1}^{N-1}(x_{i}y_{i+1}-x_{i+1}y_{i})+\alpha\sum_{i=1}^{N} \langle s_{i} \rangle \kappa_{i}^{2} + k_{s}\sum_{i=1}^{N_{A}} |\bm{r}_{i}-\bm{r}_{0i}|^{2}
\end{equation}
where the first term corresponds to the area of the irregular polygon of vertices $\bm{r}_{i}=(x_{i},y_{i})$, with $i=1\,2\ldots\,N$, and the third sum represents the energetic contribution of the $N_{A}$ adhesion points. $\kappa_{i}$ is the unsigned curvature at the vertex $i$: $\kappa_{i}=|\bm{t}_{i}-\bm{t}_{i-1}|/\langle s_{i}\rangle$ with $\bm{t}_{i}=(\bm{r}_{i+1}-\bm{r}_{i})/|\bm{r}_{i+1}-\bm{r}_{i}|$ the tangent vector at $i$ and $\langle s_{i} \rangle = (s_{i}+s_{i-1})/2$, with $s_{i}=|\bm{r}_{i+1}-\bm{r}_{i}|$. The discrete energy \eqref{eq:discrete-energy} was minimized using a standard conjugate gradient algorithm. Using \eqref{eq:discrete-energy} allows a direct comparison between simulations and the analytical results presented in the previous sections. However, for very large substrate stiffness, the discrete curve develops self-intersections and the energy becomes ill-defined. In this regime, it is more convenient to approximate the cell as a simplicial complex consisting of mesh $M$ of equilateral triangles. The edges of the triangles can then be treated as elastic springs of zero rest-length, so that the total energy of the mesh is given by:
\begin{equation}\label{eq:spring-energy}
E_{2} = \Sigma\sum_{e\in M}|e|^{2}+\alpha \sum_{v\in \partial M}\langle s_{v} \rangle\,\kappa_{v}^{2} + k_{s}\sum_{i=1}^{N_{A}} |\bm{r}_{i}-\bm{r}_{0i}|^{2}
\end{equation}
where $v$ and $e$ represent respectively the vertices and the edges of the mesh and $\Sigma$ is a spring constant. If the triangles in the mesh are equilateral, this yields a discrete approximation of the interfacial energy $\sigma A$, with the spring stiffness proportional to the surface tension: i.e. $\sigma\approx 4\Sigma\sqrt{3}/(2-B/E)$, where $B/E$ is the ratio between the number of boundary edges $B$ and the total number of edges $E$ of the triangular mesh \cite{Giomi2012}.

\section*{Appendix B $-$ Kinks, cusps and protrusions}

\begin{figure}
\centering
\includegraphics[width=1\columnwidth]{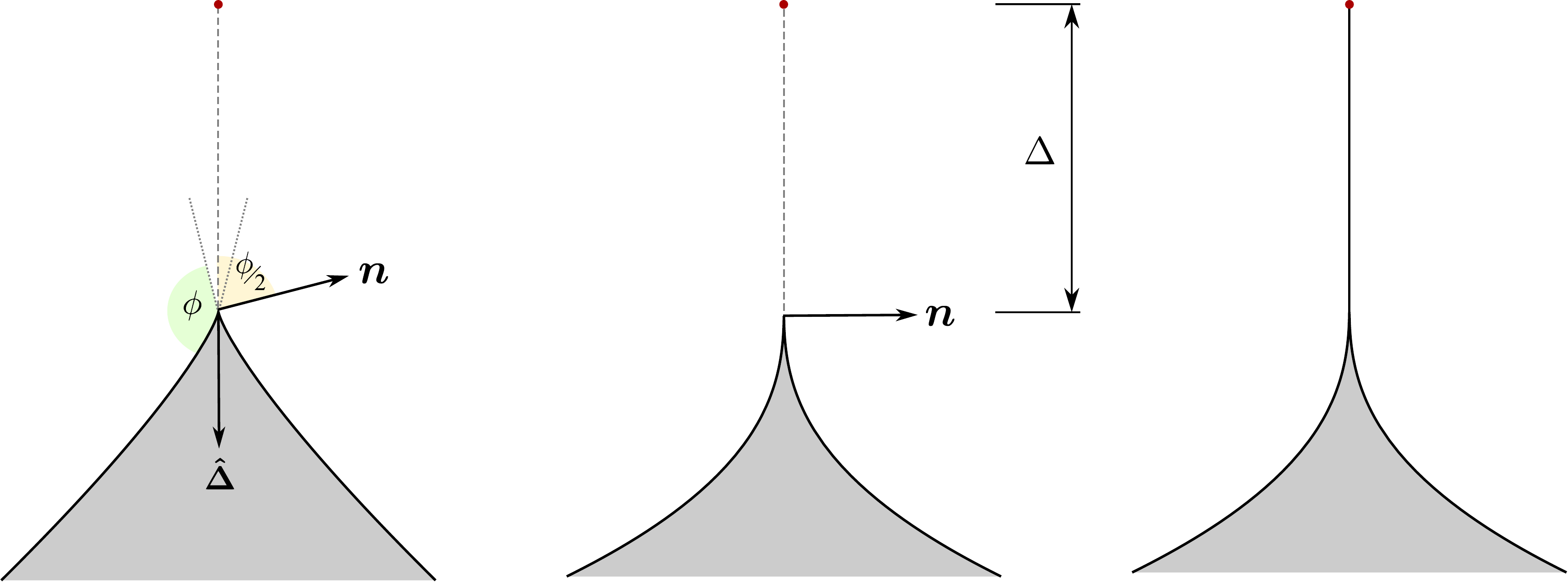}
\caption{\label{fig:singular}Example of singular points: kink (left), cusps (center), protrusion (right). The red dot indicated the adhesion point rest position $\bm{a}$, while $\bm{\hat{\Delta}}=(\bm{r}-\bm{r}_{0})/|\bm{r}-\bm{r}_{0}|$. For a cusp $\bm{n}\cdot\bm{\Delta}=0$, while for a protrusion, the normal vector is undefined at the point of adhesion.}	
\end{figure}

We present here some additional mathematical aspects on the occurrence of cusps and formation of protrusions in the large contractility and stiffness regime. In particular we show that the shape consisting of $N_{A}$ cusps that extend until the adhesion rest point through a set of straight protrusions, is the only regular convex $N_{A}$-fold star shape to be mechanically stable within the Contractile Film Model. 

A kink is a singular point on a curve where the tangent vector switches discontinuously between two orientations (Fig. \ref{fig:singular}, left). The magnitude of the discontinuity can be measured from the external angle $\phi$. A cusp, is a kink with $\phi=\pi$, so that the tangent vector switches between equal and opposite orientation (rotation by a larger angle would give rise to self-intersections). In the case of a simple closed curve with kinks, the theorem of turning tangents can be reformulated as follows:
\begin{equation}\label{eq:turning-kinks}
\oint ds\,\kappa + \sum_{i}\phi_{i} = 2\pi\;,
\end{equation}
where the summation runs over all the kinks. In the case of a convex polygon, for instance, $\kappa=0$ and \eqref{eq:turning-kinks} asserts that the sum of the external angle of a polygon is equal to $2\pi$. In a convex $N_{A}$-fold star, the external angle is bounded in the range $\phi\in[2\pi/N_{A},\pi]$, where $\phi=2\pi/N_{A}$ corresponds to a regular polygon. As described in the main text, Euler-Langrange equation for cellular force-balance is given by:
\begin{equation}\label{eq:euler-lagrange-bis}
\alpha (2\kappa''+\kappa^{3})-\sigma + 2k_{s}\sum_{i=1}^{N_{A}}\delta(s-s_{i})(\bm{r}-\bm{r}_0)\cdot\bm{n} = 0\;.
\end{equation}
Let $s_{1}=0$ be the position of a generic adhesion point. Then, integrating Eq. \eqref{eq:euler-lagrange-bis} in the range $s\in[-\epsilon,\epsilon]$ and taking the limit $\epsilon\rightarrow 0$ yields:
\begin{equation}
2\alpha \kappa'(0) + k_{s}\,\bm{n}(0)\cdot\bm{\Delta}(0) = 0\;,
\end{equation}
which expresses that the elastic restoring force originating in the boundary must balance the body force $k_{s}\,\bm{n}\cdot\bm{\Delta}$ due to the adhesion spring. For a kink, as that shown on the left of Fig. \ref{fig:singular}, $\bm{n}\cdot\bm{\Delta}=\Delta\cos(\pi-\phi/2)=-\Delta\cos(\phi/2)$, force balance gives us:
\begin{equation}\label{eq:kink}
\kappa'(0) = \frac{k_{s}}{2\alpha}\,\Delta\cos(\phi/2)\;,
\end{equation}
Now, in a configuration consisting of a regular convex $N_{A}$-fold star, the signed curvature is everywhere negative and has {\em single} minimum at the midpoint between kinks. The latter property implies $\kappa'(0)<0$, which however contradicts Eq. \eqref{eq:kink} being the right-hand side always positive for any positive value of $k_{s}$, $\alpha$ and $\Delta$. From this we conclude that such a configuration cannot be a possible equilibrium shape.

In the case of a cusp, $\bm{n}\cdot\bm{\Delta}=0$ and the adhesion force is all exerted along the tangent direction, hence $\kappa'(0)=0$. In Appendix C, however, we show that Eq. \eqref{eq:euler-lagrange-bis} has no solution with $\kappa'(0)=0$ that satisfies \eqref{eq:turning-kinks} with $\phi=\pi$. The only case left is then that illustrated on the right of Fig. \ref{fig:singular}, in which the cusp extends through a straight protrusion until the adhesion rest position, so that $\Delta=0$. In this configuration, the adhesion force exerted by the substrate is zero and so is the elastic force acting in the protrusion, this being straight. In other words, the cell is pinned at the adhesion rest position while the elastic force is zero. The case in which the protrusion has zero length, is a special instance of this scenario from which one can construct all the shapes having nonzero protrusion length by mean of a {\em similarity} transformation, described in the main text.

\section*{Appendix C $-$ Solution of the nonlinear Elastica equation}

In this section we give and exact analytical expression of the general solution of the equation:
\begin{equation}\label{eq:nonlinear}
\alpha(2\kappa''+\kappa^{3})-\sigma = 0\;,\qquad\qquad \kappa(0)=\kappa(L)=\kappa_{0}\;,
\end{equation}
and prove what asserted in Appendix B that a cusp as that shown in Fig. \ref{fig:singular} (center) cannot exist. As a starting point let us make the equation dimensionless by taking $t=s/L$ and $\hat{\kappa}=L\kappa$ and $\hat{\sigma}=\sigma L^{3}/\alpha$. We have then:
\begin{equation}
2\hat{\kappa}''+\hat{\kappa}^{3}-\hat{\sigma}=0\,,\qquad\qquad \hat{\kappa}(0)=\hat{\kappa}(1)=\hat{\kappa}_{0}
\end{equation}
where the prime now stands for a differentiation with respect to $t$. Without loss of generality, we can chose $t=0$ as the point where the derivative of $\hat{\kappa}$ vanishes. For the previously mentioned cusp, this point will be in particular a point of adhesion and $t\in[0,1]$. Otherwise, this will be identified as the mid point between adhesions upon translating $t\rightarrow t-1/2$. Then, integrating the equation with respect to $\hat{\kappa}$ and using the fact that $\hat{\kappa}(0)=\hat{\kappa}_{0}$ and $\hat{\kappa}'(0)=0$, we obtain:
\begin{equation}
(\hat{\kappa}')^{2} + \tfrac{1}{4}(\hat{\kappa}^{4}-\hat{\kappa}_{0}^{4})-\hat{\sigma}(\hat{\kappa}-\hat{\kappa}_{0}) = 0
\end{equation}
Introducing the new variable $y=1/(\hat{\kappa}_{0}-\hat{\kappa})$ we can reduce the order on the nonlinearity by one unit:
\begin{equation}
(y')^{2}=(\hat{\kappa}_{0}^{3}-\sigma)y^{3}-\tfrac{3}{2}\hat{\kappa}_{0}^{2}y^{2}+\hat{\kappa}_{0}y-\tfrac{1}{4}	
\end{equation}
This equation is of the form:
\begin{equation}
(y')^{2} = P(y) = h^{2}(y-a)(y-b)(y-c)	
\end{equation}
with $h^{2}=\hat{\kappa}_{0}^{3}-\sigma$ and $a$, $b$ and $c$ the roots of the cubic polynomial $P(y)$, and is suitable to be solved in terms of elliptic functions. Now, one can verify that $P(y)$ has always a single real root, $y=\alpha$ and a pair of complex conjugate roots $y=\beta\pm i\gamma$ for any physical value of $\hat{\kappa}_{0}$ and $\hat{\sigma}_{0}$. Thus:
\begin{equation}
P(y) = h^{2}(y-\alpha)[(y-\beta)^{2}+\gamma^{2}]
\end{equation}
which allows us to calculate the elliptic integral \cite{Greenhill1892}:
\begin{equation}\label{eq:y}
t = \int_{y}^{\infty} \frac{dy}{\sqrt{P(y)}} = \omega^{-1}\cn^{-1}\left(\frac{y-z_{1}}{y-z_{2}},m\right)
\end{equation}
where $z_{1}$ and $z_{2}$ are the roots of the quadratic equations:
\begin{equation}\label{eq:quadratic}
z^{2}-2\alpha z + 2\alpha\beta-(\beta^{2}+\gamma^{2}) = 0	
\end{equation}
and $\omega$ and $m$ is given by:
\begin{equation}
\omega^{2}=\frac{h(z_{1}-z_{2})}{2}\,,\qquad m^{2} = \frac{\beta-z_{2}}{z_{1}-z_{2}}	
\end{equation}
Finally, solving \eqref{eq:y} for $y$ and going back to our original variables, we have:
\begin{equation}\label{eq:kappasol}
\hat{\kappa}=\hat{\kappa}_{0}-\frac{1-\cn(\omega t,m)}{z_{1}-z_{2}\cn(\omega t,m)}	
\end{equation}
\begin{figure}
\centering
\includegraphics[width=1\columnwidth]{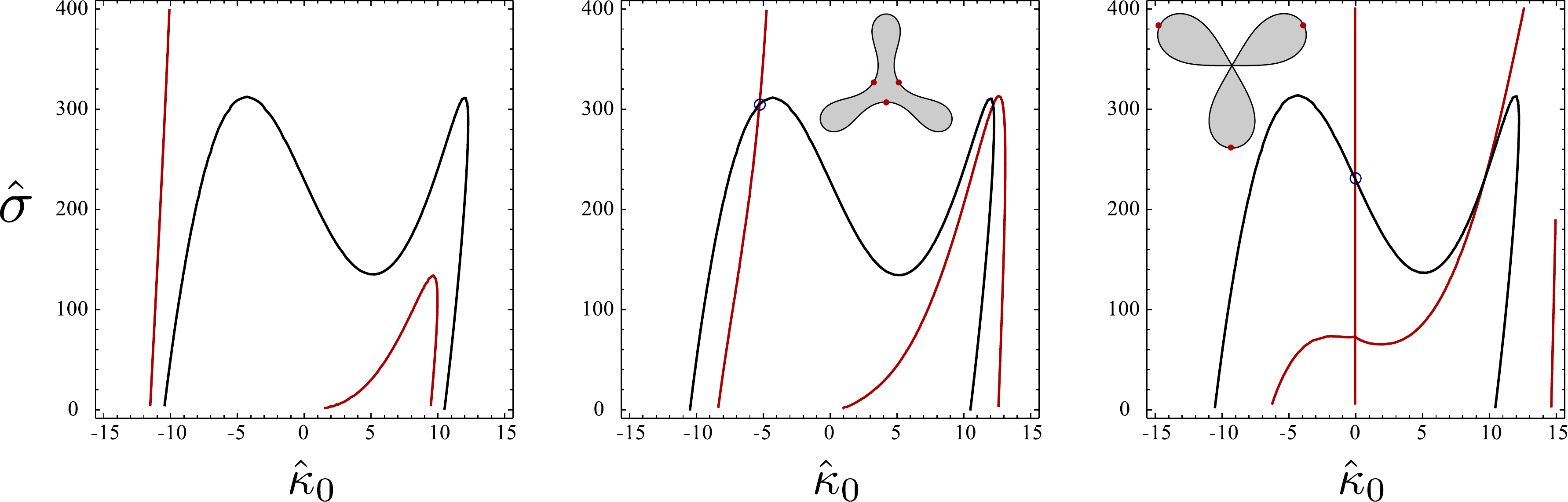}
\caption{\label{fig:roots}The curve obtained by solving Eqs. \eqref{eq:period} (black) and \eqref{eq:closure} (red). The left panel corresponds to the cusp shown in Fig. \ref{fig:singular} (left). The absence of intersections between the curves implies that such a configuration is not mechanically stable.}	
\end{figure}
In Eqs. \eqref{eq:y} and \eqref{eq:kappasol} we use the standard notation for Jacobi elliptic functions \cite{Davis2010}. Namely, given the incomplete elliptic integral of the first kind:
\[
u = F(\phi,m) = \int_{0}^{\phi} \frac{dt}{\sqrt{1-m^{2}\sin^{2}t}}
\]
with $0<m^{2}<1$, the elliptic modulus, then $\phi$ is the Jacobi amplitude: $\phi=\am(u,m)$ and $\cn(u,m)=\cos\phi$.

The expression \eqref{eq:kappasol}, satisfies by construction the boundary conditions $\hat{\kappa}(0)=\hat{\kappa}_{0}$ and $\hat{\kappa}'(0)=0$. In order for it to be a legitimate solution of the problem, we further need it to be periodic, so that $\hat{\kappa}(0)=\hat{\kappa}(1)$ and to satisfy the theorem of turning tangents \eqref{eq:turning-kinks}. Periodicity can be easily implemented by recalling that $\cn(x+4K(m))=\cn(x)$, where $K(m)=F(\frac{\pi}{2},m)$ is the complete elliptic integral of first kind. This results in the following condition for the frequency: $\omega=4K(m)$, or more explicitly:
\begin{equation}\label{eq:period}
h(z_{1}-z_{2}) = 32K^{2}(m)	
\end{equation}
From the theorem of turning tangents applied to the case of a simple closed curve with $N_{A}$ cups, we obtain :
\begin{multline}\label{eq:closure}
\hat{\kappa}_{0}-\pi\left(\frac{2-N_{A}}{N_{A}}\right)\\ =\frac{4}{z_{1}z_{2}}\left[z_{1}K(m)-\frac{z_{1}-z_{2}}{\sqrt{1-m^{2}}}\,\Pi\left(\frac{z_{2}^{2}}{z_{1}^{2}}\,\bigg|\,\frac{m^{2}}{m^{2}-1}\right)\right]
\end{multline}
where $\Pi$ is the complete elliptic integral of third kind:
\begin{equation}
\Pi(n\,|\,m) = \int_{0}^{\frac{\pi}{2}}\frac{dt}{(1-n\sin^{2}t)\sqrt{1-m\sin^{2}t}}
\end{equation}
Solving simultaneously the transcendental equations \eqref{eq:period} and \eqref{eq:closure}, the quadratic equation \eqref{eq:quadratic} and its associated cubic, allows to calculate $\hat{\kappa}_{0}$ and $\hat{\sigma}$, from which one can obtain $\kappa_{0}$ and $L$. The solution is then complete.
With this machinery in hand we can now answer the original question. Is there any solution corresponding to the cusp in the center of Fig. \ref{fig:singular}, with $\hat{\kappa}'(0)=0$ and, say, $N_{A}=3$, such that $\int_{0}^{1}dt\,\hat{\kappa}=-\pi/3$? A numerical solution of Eqs. \eqref{eq:period} (black) and \eqref{eq:closure} (red) is plotted in Fig. \ref{fig:roots} as a function of $\hat{\kappa}_{0}$ and $\hat{\sigma}$. The curves never intersect in the physical range of $\hat{\kappa}_{0}$ and $\hat{\sigma}$, thus such a solution does not exist.

\bibliography{cell-shape}

\begin{thebibliography}{49}%
\makeatletter
\providecommand \@ifxundefined [1]{%
 \@ifx{#1\undefined}
}%
\providecommand \@ifnum [1]{%
 \ifnum #1\expandafter \@firstoftwo
 \else \expandafter \@secondoftwo
 \fi
}%
\providecommand \@ifx [1]{%
 \ifx #1\expandafter \@firstoftwo
 \else \expandafter \@secondoftwo
 \fi
}%
\providecommand \natexlab [1]{#1}%
\providecommand \enquote  [1]{``#1''}%
\providecommand \bibnamefont  [1]{#1}%
\providecommand \bibfnamefont [1]{#1}%
\providecommand \citenamefont [1]{#1}%
\providecommand \href@noop [0]{\@secondoftwo}%
\providecommand \href [0]{\begingroup \@sanitize@url \@href}%
\providecommand \@href[1]{\@@startlink{#1}\@@href}%
\providecommand \@@href[1]{\endgroup#1\@@endlink}%
\providecommand \@sanitize@url [0]{\catcode `\\12\catcode `\$12\catcode
  `\&12\catcode `\#12\catcode `\^12\catcode `\_12\catcode `\%12\relax}%
\providecommand \@@startlink[1]{}%
\providecommand \@@endlink[0]{}%
\providecommand \url  [0]{\begingroup\@sanitize@url \@url }%
\providecommand \@url [1]{\endgroup\@href {#1}{\urlprefix }}%
\providecommand \urlprefix  [0]{URL }%
\providecommand \Eprint [0]{\href }%
\@ifxundefined \urlstyle {%
  \providecommand \doi  [0]{\begingroup \@sanitize@url \@doi}%
  \providecommand \@doi [1]{\endgroup \@@startlink {\doibase
  #1}doi:\discretionary {}{}{}#1\@@endlink }%
}{%
  \providecommand \doi  [0]{doi:\discretionary{}{}{}\begingroup
  \urlstyle{rm}\Url }%
}%
\providecommand \doibase [0]{http://dx.doi.org/}%
\providecommand \Doi [0]{\begingroup \@sanitize@url \@Doi }%
\providecommand \@Doi  [1]{\endgroup\@@startlink{\doibase#1}\@@Doi}%
\providecommand \@@Doi [1]{#1\@@endlink}%
\providecommand \selectlanguage [0]{\@gobble}%
\providecommand \bibinfo  [0]{\@secondoftwo}%
\providecommand \bibfield  [0]{\@secondoftwo}%
\providecommand \translation [1]{[#1]}%
\providecommand \BibitemOpen [0]{}%
\providecommand \bibitemStop [0]{}%
\providecommand \bibitemNoStop [0]{.\EOS\space}%
\providecommand \EOS [0]{\spacefactor3000\relax}%
\providecommand \BibitemShut  [1]{\csname bibitem#1\endcsname}%
\bibitem [{\citenamefont {Discher}\ \emph {et~al.}(2005)\citenamefont
  {Discher}, \citenamefont {Janmey},\ and\ \citenamefont {Wang}}]{Discher2005}%
  \BibitemOpen
  \bibfield  {author} {\bibinfo {author} {\bibfnamefont {D.}~\bibnamefont
  {Discher}}, \bibinfo {author} {\bibfnamefont {P.}~\bibnamefont {Janmey}}, \
  and\ \bibinfo {author} {\bibfnamefont {Y.}~\bibnamefont {Wang}},\ }\href@noop
  {} {\bibfield  {journal} {\bibinfo  {journal} {Science},\ }\textbf {\bibinfo
  {volume} {310}},\ \bibinfo {pages} {1139} (\bibinfo {year}
  {2005})}\BibitemShut {NoStop}%
\bibitem [{\citenamefont {Yeung}\ \emph {et~al.}(2005)\citenamefont {Yeung},
  \citenamefont {Georges}, \citenamefont {Flanagan}, \citenamefont {Marg},
  \citenamefont {Ortiz}, \citenamefont {Funaki}, \citenamefont {Zahir},
  \citenamefont {Ming}, \citenamefont {Weaver},\ and\ \citenamefont
  {Janmey}}]{Yeung2005}%
  \BibitemOpen
  \bibfield  {author} {\bibinfo {author} {\bibfnamefont {T.}~\bibnamefont
  {Yeung}}, \bibinfo {author} {\bibfnamefont {P.}~\bibnamefont {Georges}},
  \bibinfo {author} {\bibfnamefont {L.}~\bibnamefont {Flanagan}}, \bibinfo
  {author} {\bibfnamefont {B.}~\bibnamefont {Marg}}, \bibinfo {author}
  {\bibfnamefont {M.}~\bibnamefont {Ortiz}}, \bibinfo {author} {\bibfnamefont
  {M.}~\bibnamefont {Funaki}}, \bibinfo {author} {\bibfnamefont
  {N.}~\bibnamefont {Zahir}}, \bibinfo {author} {\bibfnamefont
  {W.}~\bibnamefont {Ming}}, \bibinfo {author} {\bibfnamefont {V.}~\bibnamefont
  {Weaver}}, \ and\ \bibinfo {author} {\bibfnamefont {P.}~\bibnamefont
  {Janmey}},\ }\href@noop {} {\bibfield  {journal} {\bibinfo  {journal} {Cell
  Motil Cytoskel},\ }\textbf {\bibinfo {volume} {60}},\ \bibinfo {pages} {24}
  (\bibinfo {year} {2005})}\BibitemShut {NoStop}%
\bibitem [{\citenamefont {Chopra}\ \emph {et~al.}(2011)\citenamefont {Chopra},
  \citenamefont {Tabdanov}, \citenamefont {Patel}, \citenamefont {Janmey},\
  and\ \citenamefont {Kresh}}]{Chopra2011}%
  \BibitemOpen
  \bibfield  {author} {\bibinfo {author} {\bibfnamefont {A.}~\bibnamefont
  {Chopra}}, \bibinfo {author} {\bibfnamefont {E.}~\bibnamefont {Tabdanov}},
  \bibinfo {author} {\bibfnamefont {H.}~\bibnamefont {Patel}}, \bibinfo
  {author} {\bibfnamefont {P.}~\bibnamefont {Janmey}}, \ and\ \bibinfo {author}
  {\bibfnamefont {J.}~\bibnamefont {Kresh}},\ }\href@noop {} {\bibfield
  {journal} {\bibinfo  {journal} {Am J Physiol-Heart C},\ }\textbf {\bibinfo
  {volume} {300}},\ \bibinfo {pages} {H1252} (\bibinfo {year}
  {2011})}\BibitemShut {NoStop}%
\bibitem [{\citenamefont {Chen}\ \emph {et~al.}(1997)\citenamefont {Chen},
  \citenamefont {Mrksich}, \citenamefont {Huang}, \citenamefont {Whitesides},\
  and\ \citenamefont {Ingber}}]{Chen1997}%
  \BibitemOpen
  \bibfield  {author} {\bibinfo {author} {\bibfnamefont {C.}~\bibnamefont
  {Chen}}, \bibinfo {author} {\bibfnamefont {M.}~\bibnamefont {Mrksich}},
  \bibinfo {author} {\bibfnamefont {S.}~\bibnamefont {Huang}}, \bibinfo
  {author} {\bibfnamefont {G.}~\bibnamefont {Whitesides}}, \ and\ \bibinfo
  {author} {\bibfnamefont {D.}~\bibnamefont {Ingber}},\ }\href@noop {}
  {\bibfield  {journal} {\bibinfo  {journal} {Science},\ }\textbf {\bibinfo
  {volume} {276}},\ \bibinfo {pages} {1425} (\bibinfo {year}
  {1997})}\BibitemShut {NoStop}%
\bibitem [{\citenamefont {Galbraith}\ and\ \citenamefont
  {Sheetz}(1998)}]{Galbraith1998}%
  \BibitemOpen
  \bibfield  {author} {\bibinfo {author} {\bibfnamefont {C.}~\bibnamefont
  {Galbraith}}\ and\ \bibinfo {author} {\bibfnamefont {M.}~\bibnamefont
  {Sheetz}},\ }\href@noop {} {\bibfield  {journal} {\bibinfo  {journal} {Curr
  Opin Cell Biol},\ }\textbf {\bibinfo {volume} {10}},\ \bibinfo {pages} {566}
  (\bibinfo {year} {1998})}\BibitemShut {NoStop}%
\bibitem [{\citenamefont {Riveline}\ \emph {et~al.}(2001)\citenamefont
  {Riveline}, \citenamefont {Zamir}, \citenamefont {Balaban}, \citenamefont
  {Schwarz}, \citenamefont {Ishizaki}, \citenamefont {Narumiya}, \citenamefont
  {Kam}, \citenamefont {Geiger},\ and\ \citenamefont
  {Bershadsky}}]{Riveline2001}%
  \BibitemOpen
  \bibfield  {author} {\bibinfo {author} {\bibfnamefont {D.}~\bibnamefont
  {Riveline}}, \bibinfo {author} {\bibfnamefont {E.}~\bibnamefont {Zamir}},
  \bibinfo {author} {\bibfnamefont {N.}~\bibnamefont {Balaban}}, \bibinfo
  {author} {\bibfnamefont {U.}~\bibnamefont {Schwarz}}, \bibinfo {author}
  {\bibfnamefont {T.}~\bibnamefont {Ishizaki}}, \bibinfo {author}
  {\bibfnamefont {S.}~\bibnamefont {Narumiya}}, \bibinfo {author}
  {\bibfnamefont {Z.}~\bibnamefont {Kam}}, \bibinfo {author} {\bibfnamefont
  {B.}~\bibnamefont {Geiger}}, \ and\ \bibinfo {author} {\bibfnamefont
  {A.}~\bibnamefont {Bershadsky}},\ }\href@noop {} {\bibfield  {journal}
  {\bibinfo  {journal} {Science's STKE},\ }\textbf {\bibinfo {volume} {153}},\
  \bibinfo {pages} {1175} (\bibinfo {year} {2001})}\BibitemShut {NoStop}%
\bibitem [{\citenamefont {Bar-Ziv}\ \emph {et~al.}(1999)\citenamefont
  {Bar-Ziv}, \citenamefont {Tlusty}, \citenamefont {Moses}, \citenamefont
  {Safran},\ and\ \citenamefont {Bershadsky}}]{Barziv1999}%
  \BibitemOpen
  \bibfield  {author} {\bibinfo {author} {\bibfnamefont {R.}~\bibnamefont
  {Bar-Ziv}}, \bibinfo {author} {\bibfnamefont {T.}~\bibnamefont {Tlusty}},
  \bibinfo {author} {\bibfnamefont {E.}~\bibnamefont {Moses}}, \bibinfo
  {author} {\bibfnamefont {S.}~\bibnamefont {Safran}}, \ and\ \bibinfo {author}
  {\bibfnamefont {A.}~\bibnamefont {Bershadsky}},\ }\href@noop {} {\bibfield
  {journal} {\bibinfo  {journal} {Proc Natl Acad Sci USA},\ }\textbf {\bibinfo
  {volume} {96}},\ \bibinfo {pages} {10140} (\bibinfo {year}
  {1999})}\BibitemShut {NoStop}%
\bibitem [{\citenamefont {Asano}\ \emph {et~al.}(2009)\citenamefont {Asano},
  \citenamefont {Jim{\'e}nez-Dalmaroni}, \citenamefont {Liverpool},
  \citenamefont {Marchetti}, \citenamefont {Giomi}, \citenamefont {Kiger},
  \citenamefont {Duke},\ and\ \citenamefont {Baum}}]{Asano2009}%
  \BibitemOpen
  \bibfield  {author} {\bibinfo {author} {\bibfnamefont {Y.}~\bibnamefont
  {Asano}}, \bibinfo {author} {\bibfnamefont {A.}~\bibnamefont
  {Jim{\'e}nez-Dalmaroni}}, \bibinfo {author} {\bibfnamefont {T.}~\bibnamefont
  {Liverpool}}, \bibinfo {author} {\bibfnamefont {M.}~\bibnamefont
  {Marchetti}}, \bibinfo {author} {\bibfnamefont {L.}~\bibnamefont {Giomi}},
  \bibinfo {author} {\bibfnamefont {A.}~\bibnamefont {Kiger}}, \bibinfo
  {author} {\bibfnamefont {T.}~\bibnamefont {Duke}}, \ and\ \bibinfo {author}
  {\bibfnamefont {B.}~\bibnamefont {Baum}},\ }\href@noop {} {\bibfield
  {journal} {\bibinfo  {journal} {HFSP journal},\ }\textbf {\bibinfo {volume}
  {3}},\ \bibinfo {pages} {194} (\bibinfo {year} {2009})}\BibitemShut {NoStop}%
\bibitem [{\citenamefont {Th{\'e}ry}\ \emph {et~al.}(2006)\citenamefont
  {Th{\'e}ry}, \citenamefont {P{\'e}pin}, \citenamefont {Dressaire},
  \citenamefont {Chen},\ and\ \citenamefont {Bornens}}]{Thery2006}%
  \BibitemOpen
  \bibfield  {author} {\bibinfo {author} {\bibfnamefont {M.}~\bibnamefont
  {Th{\'e}ry}}, \bibinfo {author} {\bibfnamefont {A.}~\bibnamefont
  {P{\'e}pin}}, \bibinfo {author} {\bibfnamefont {E.}~\bibnamefont
  {Dressaire}}, \bibinfo {author} {\bibfnamefont {Y.}~\bibnamefont {Chen}}, \
  and\ \bibinfo {author} {\bibfnamefont {M.}~\bibnamefont {Bornens}},\
  }\href@noop {} {\bibfield  {journal} {\bibinfo  {journal} {Cell Motil
  Cytoskel},\ }\textbf {\bibinfo {volume} {63}},\ \bibinfo {pages} {341}
  (\bibinfo {year} {2006})}\BibitemShut {NoStop}%
\bibitem [{\citenamefont {Pelham}\ and\ \citenamefont
  {Wang}(1997)}]{Pelham1997}%
  \BibitemOpen
  \bibfield  {author} {\bibinfo {author} {\bibfnamefont {R.}~\bibnamefont
  {Pelham}}\ and\ \bibinfo {author} {\bibfnamefont {Y.}~\bibnamefont {Wang}},\
  }\href@noop {} {\bibfield  {journal} {\bibinfo  {journal} {Proc Natl Acad Sci
  USA},\ }\textbf {\bibinfo {volume} {94}},\ \bibinfo {pages} {13661} (\bibinfo
  {year} {1997})}\BibitemShut {NoStop}%
\bibitem [{\citenamefont {Du~Roure}\ \emph {et~al.}(2005)\citenamefont
  {Du~Roure}, \citenamefont {Saez}, \citenamefont {Buguin}, \citenamefont
  {Austin}, \citenamefont {Chavrier}, \citenamefont {Siberzan},\ and\
  \citenamefont {Ladoux}}]{duroure2005}%
  \BibitemOpen
  \bibfield  {author} {\bibinfo {author} {\bibfnamefont {O.}~\bibnamefont
  {Du~Roure}}, \bibinfo {author} {\bibfnamefont {A.}~\bibnamefont {Saez}},
  \bibinfo {author} {\bibfnamefont {A.}~\bibnamefont {Buguin}}, \bibinfo
  {author} {\bibfnamefont {R.}~\bibnamefont {Austin}}, \bibinfo {author}
  {\bibfnamefont {P.}~\bibnamefont {Chavrier}}, \bibinfo {author}
  {\bibfnamefont {P.}~\bibnamefont {Siberzan}}, \ and\ \bibinfo {author}
  {\bibfnamefont {B.}~\bibnamefont {Ladoux}},\ }\href@noop {} {\bibfield
  {journal} {\bibinfo  {journal} {Proc Natl Acad Sci USA},\ }\textbf {\bibinfo
  {volume} {102}},\ \bibinfo {pages} {2390} (\bibinfo {year}
  {2005})}\BibitemShut {NoStop}%
\bibitem [{\citenamefont {Lecuit}\ and\ \citenamefont
  {Lenne}(2007)}]{Lecuit2007}%
  \BibitemOpen
  \bibfield  {author} {\bibinfo {author} {\bibfnamefont {T.}~\bibnamefont
  {Lecuit}}\ and\ \bibinfo {author} {\bibfnamefont {P.}~\bibnamefont {Lenne}},\
  }\href@noop {} {\bibfield  {journal} {\bibinfo  {journal} {Nat Rev Mol Cell
  Bio},\ }\textbf {\bibinfo {volume} {8}},\ \bibinfo {pages} {633} (\bibinfo
  {year} {2007})}\BibitemShut {NoStop}%
\bibitem [{\citenamefont {Mader}\ \emph {et~al.}(2007)\citenamefont {Mader},
  \citenamefont {Hinchcliffe},\ and\ \citenamefont {Wang}}]{Mader2007}%
  \BibitemOpen
  \bibfield  {author} {\bibinfo {author} {\bibfnamefont {C.}~\bibnamefont
  {Mader}}, \bibinfo {author} {\bibfnamefont {E.}~\bibnamefont {Hinchcliffe}},
  \ and\ \bibinfo {author} {\bibfnamefont {Y.}~\bibnamefont {Wang}},\
  }\href@noop {} {\bibfield  {journal} {\bibinfo  {journal} {Soft Matter},\
  }\textbf {\bibinfo {volume} {3}},\ \bibinfo {pages} {357} (\bibinfo {year}
  {2007})}\BibitemShut {NoStop}%
\bibitem [{\citenamefont {Buchsbaum}(2007)}]{Buchsbaum2007}%
  \BibitemOpen
  \bibfield  {author} {\bibinfo {author} {\bibfnamefont {R.}~\bibnamefont
  {Buchsbaum}},\ }\href@noop {} {\bibfield  {journal} {\bibinfo  {journal} {J
  Cell Sci},\ }\textbf {\bibinfo {volume} {120}},\ \bibinfo {pages} {1149}
  (\bibinfo {year} {2007})}\BibitemShut {NoStop}%
\bibitem [{\citenamefont {Wozniak}\ \emph {et~al.}(2004)\citenamefont
  {Wozniak}, \citenamefont {Modzelewska}, \citenamefont {Kwong},\ and\
  \citenamefont {Keely}}]{Wozniak2004}%
  \BibitemOpen
  \bibfield  {author} {\bibinfo {author} {\bibfnamefont {M.}~\bibnamefont
  {Wozniak}}, \bibinfo {author} {\bibfnamefont {K.}~\bibnamefont
  {Modzelewska}}, \bibinfo {author} {\bibfnamefont {L.}~\bibnamefont {Kwong}},
  \ and\ \bibinfo {author} {\bibfnamefont {P.}~\bibnamefont {Keely}},\
  }\href@noop {} {\bibfield  {journal} {\bibinfo  {journal} {BBA-Mol Cell
  Res},\ }\textbf {\bibinfo {volume} {1692}},\ \bibinfo {pages} {103} (\bibinfo
  {year} {2004})}\BibitemShut {NoStop}%
\bibitem [{\citenamefont {Howard}(2001)}]{Howard2001}%
  \BibitemOpen
  \bibfield  {author} {\bibinfo {author} {\bibfnamefont {J.}~\bibnamefont
  {Howard}},\ }\href@noop {} {\emph {\bibinfo {title} {Mechanics of motor
  proteins and the cytoskeleton}}}\ (\bibinfo  {publisher} {Sinauer Associates
  Sunderland, MA},\ \bibinfo {year} {2001})\BibitemShut {NoStop}%
\bibitem [{\citenamefont {Lemmon}\ \emph {et~al.}(2005)\citenamefont {Lemmon},
  \citenamefont {Sniadecki}, \citenamefont {Ruiz}, \citenamefont {Tan},
  \citenamefont {Romer},\ and\ \citenamefont {Chen}}]{Lemmon2005}%
  \BibitemOpen
  \bibfield  {author} {\bibinfo {author} {\bibfnamefont {C.}~\bibnamefont
  {Lemmon}}, \bibinfo {author} {\bibfnamefont {N.}~\bibnamefont {Sniadecki}},
  \bibinfo {author} {\bibfnamefont {S.}~\bibnamefont {Ruiz}}, \bibinfo {author}
  {\bibfnamefont {J.}~\bibnamefont {Tan}}, \bibinfo {author} {\bibfnamefont
  {L.}~\bibnamefont {Romer}}, \ and\ \bibinfo {author} {\bibfnamefont
  {C.}~\bibnamefont {Chen}},\ }\href@noop {} {\bibfield  {journal} {\bibinfo
  {journal} {Mechanics \& chemistry of biosystems: MCB},\ }\textbf {\bibinfo
  {volume} {2}},\ \bibinfo {pages} {1} (\bibinfo {year} {2005})}\BibitemShut
  {NoStop}%
\bibitem [{\citenamefont {Bischofs}\ \emph {et~al.}(2009)\citenamefont
  {Bischofs}, \citenamefont {Schmidt},\ and\ \citenamefont
  {Schwarz}}]{Bischofs2009}%
  \BibitemOpen
  \bibfield  {author} {\bibinfo {author} {\bibfnamefont {I.}~\bibnamefont
  {Bischofs}}, \bibinfo {author} {\bibfnamefont {S.}~\bibnamefont {Schmidt}}, \
  and\ \bibinfo {author} {\bibfnamefont {U.}~\bibnamefont {Schwarz}},\
  }\href@noop {} {\bibfield  {journal} {\bibinfo  {journal} {Phys Rev Lett},\
  }\textbf {\bibinfo {volume} {103}},\ \bibinfo {pages} {48101} (\bibinfo
  {year} {2009})}\BibitemShut {NoStop}%
\bibitem [{\citenamefont {Mertz}\ \emph {et~al.}(2012)\citenamefont {Mertz},
  \citenamefont {Banerjee}, \citenamefont {Che}, \citenamefont {German},
  \citenamefont {Xu}, \citenamefont {Hyland}, \citenamefont {Marchetti},
  \citenamefont {Horsley},\ and\ \citenamefont {Dufresne}}]{Mertz2012}%
  \BibitemOpen
  \bibfield  {author} {\bibinfo {author} {\bibfnamefont {A.}~\bibnamefont
  {Mertz}}, \bibinfo {author} {\bibfnamefont {S.}~\bibnamefont {Banerjee}},
  \bibinfo {author} {\bibfnamefont {Y.}~\bibnamefont {Che}}, \bibinfo {author}
  {\bibfnamefont {G.}~\bibnamefont {German}}, \bibinfo {author} {\bibfnamefont
  {Y.}~\bibnamefont {Xu}}, \bibinfo {author} {\bibfnamefont {C.}~\bibnamefont
  {Hyland}}, \bibinfo {author} {\bibfnamefont {M.}~\bibnamefont {Marchetti}},
  \bibinfo {author} {\bibfnamefont {V.}~\bibnamefont {Horsley}}, \ and\
  \bibinfo {author} {\bibfnamefont {E.}~\bibnamefont {Dufresne}},\ }\href@noop
  {} {\bibfield  {journal} {\bibinfo  {journal} {Phys Rev Lett},\ }\textbf
  {\bibinfo {volume} {108}},\ \bibinfo {pages} {198101} (\bibinfo {year}
  {2012})}\BibitemShut {NoStop}%
\bibitem [{\citenamefont {Bischofs}\ \emph {et~al.}(2008)\citenamefont
  {Bischofs}, \citenamefont {Klein}, \citenamefont {Lehnert}, \citenamefont
  {Bastmeyer},\ and\ \citenamefont {Schwarz}}]{Bischofs2008}%
  \BibitemOpen
  \bibfield  {author} {\bibinfo {author} {\bibfnamefont {I.}~\bibnamefont
  {Bischofs}}, \bibinfo {author} {\bibfnamefont {F.}~\bibnamefont {Klein}},
  \bibinfo {author} {\bibfnamefont {D.}~\bibnamefont {Lehnert}}, \bibinfo
  {author} {\bibfnamefont {M.}~\bibnamefont {Bastmeyer}}, \ and\ \bibinfo
  {author} {\bibfnamefont {U.}~\bibnamefont {Schwarz}},\ }\href@noop {}
  {\bibfield  {journal} {\bibinfo  {journal} {Biophys J},\ }\textbf {\bibinfo
  {volume} {95}},\ \bibinfo {pages} {3488} (\bibinfo {year}
  {2008})}\BibitemShut {NoStop}%
\bibitem [{\citenamefont {Mumford}(1993)}]{Mumford1993}%
  \BibitemOpen
  \bibfield  {author} {\bibinfo {author} {\bibfnamefont {D.}~\bibnamefont
  {Mumford}},\ }in\ \href@noop {} {\emph {\bibinfo {booktitle} {Algebraic
  geometry and its applications}}},\ \bibinfo {editor} {edited by\ \bibinfo
  {editor} {\bibfnamefont {C.}~\bibnamefont {Bajaj}}}\ (\bibinfo  {publisher}
  {Springer-Verlag, New York},\ \bibinfo {year} {1993})\ pp.\ \bibinfo {pages}
  {507--518}\BibitemShut {NoStop}%
\bibitem [{\citenamefont {Giomi}\ and\ \citenamefont
  {Mahadevan}(2012)}]{Giomi2012}%
  \BibitemOpen
  \bibfield  {author} {\bibinfo {author} {\bibfnamefont {L.}~\bibnamefont
  {Giomi}}\ and\ \bibinfo {author} {\bibfnamefont {L.}~\bibnamefont
  {Mahadevan}},\ }\href@noop {} {\bibfield  {journal} {\bibinfo  {journal}
  {Proc R Soc A},\ }\textbf {\bibinfo {volume} {468}},\ \bibinfo {pages} {1851}
  (\bibinfo {year} {2012})}\BibitemShut {NoStop}%
\bibitem [{\citenamefont {Deshpande}\ \emph {et~al.}(2006)\citenamefont
  {Deshpande}, \citenamefont {McMeeking},\ and\ \citenamefont
  {Evans}}]{Deshpande2006}%
  \BibitemOpen
  \bibfield  {author} {\bibinfo {author} {\bibfnamefont {V.~S.}\ \bibnamefont
  {Deshpande}}, \bibinfo {author} {\bibfnamefont {R.~M.}\ \bibnamefont
  {McMeeking}}, \ and\ \bibinfo {author} {\bibfnamefont {A.~G.}\ \bibnamefont
  {Evans}},\ }\href@noop {} {\bibfield  {journal} {\bibinfo  {journal} {Proc
  Natl Acad Sci USA},\ }\textbf {\bibinfo {volume} {103}},\ \bibinfo {pages}
  {14015} (\bibinfo {year} {2006})}\BibitemShut {NoStop}%
\bibitem [{\citenamefont {Loosli}\ \emph {et~al.}(2010)\citenamefont {Loosli},
  \citenamefont {Luginbuehl},\ and\ \citenamefont {Snedeker}}]{Loosli2010}%
  \BibitemOpen
  \bibfield  {author} {\bibinfo {author} {\bibfnamefont {Y.}~\bibnamefont
  {Loosli}}, \bibinfo {author} {\bibfnamefont {R.}~\bibnamefont {Luginbuehl}},
  \ and\ \bibinfo {author} {\bibfnamefont {J.~G.}\ \bibnamefont {Snedeker}},\
  }\href@noop {} {\bibfield  {journal} {\bibinfo  {journal} {Phil Trans R Soc
  A},\ }\textbf {\bibinfo {volume} {368}},\ \bibinfo {pages} {2629} (\bibinfo
  {year} {2010})}\BibitemShut {NoStop}%
\bibitem [{\citenamefont {Farsad}\ and\ \citenamefont
  {Vernerey}(2012)}]{Farsad2012}%
  \BibitemOpen
  \bibfield  {author} {\bibinfo {author} {\bibfnamefont {M.}~\bibnamefont
  {Farsad}}\ and\ \bibinfo {author} {\bibfnamefont {F.~J.}\ \bibnamefont
  {Vernerey}},\ }\href@noop {} {\bibfield  {journal} {\bibinfo  {journal} {Int
  J Num Meth Eng},\ }\textbf {\bibinfo {volume} {92}},\ \bibinfo {pages} {238}
  (\bibinfo {year} {2012})}\BibitemShut {NoStop}%
\bibitem [{\citenamefont {Banerjee}\ and\ \citenamefont
  {Marchetti}(2011)}]{Banerjee2011}%
  \BibitemOpen
  \bibfield  {author} {\bibinfo {author} {\bibfnamefont {S.}~\bibnamefont
  {Banerjee}}\ and\ \bibinfo {author} {\bibfnamefont {M.~C.}\ \bibnamefont
  {Marchetti}},\ }\href@noop {} {\bibfield  {journal} {\bibinfo  {journal} {EPL
  (Europhysics Letters)},\ }\textbf {\bibinfo {volume} {96}},\ \bibinfo {pages}
  {28003} (\bibinfo {year} {2011})}\BibitemShut {NoStop}%
\bibitem [{\citenamefont {Edwards}\ and\ \citenamefont
  {Schwarz}(2011)}]{Edwards2011}%
  \BibitemOpen
  \bibfield  {author} {\bibinfo {author} {\bibfnamefont {C.~M.}\ \bibnamefont
  {Edwards}}\ and\ \bibinfo {author} {\bibfnamefont {U.~S.}\ \bibnamefont
  {Schwarz}},\ }\href@noop {} {\bibfield  {journal} {\bibinfo  {journal}
  {Physical Review Letters},\ }\textbf {\bibinfo {volume} {107}},\ \bibinfo
  {pages} {128101} (\bibinfo {year} {2011})}\BibitemShut {NoStop}%
\bibitem [{\citenamefont {Torres}\ \emph {et~al.}(2012)\citenamefont {Torres},
  \citenamefont {Bischofs},\ and\ \citenamefont {Schwarz}}]{Torres2012}%
  \BibitemOpen
  \bibfield  {author} {\bibinfo {author} {\bibfnamefont {P.~G.}\ \bibnamefont
  {Torres}}, \bibinfo {author} {\bibfnamefont {I.}~\bibnamefont {Bischofs}}, \
  and\ \bibinfo {author} {\bibfnamefont {U.}~\bibnamefont {Schwarz}},\
  }\href@noop {} {\bibfield  {journal} {\bibinfo  {journal} {Physical Review
  E},\ }\textbf {\bibinfo {volume} {85}},\ \bibinfo {pages} {011913} (\bibinfo
  {year} {2012})}\BibitemShut {NoStop}%
\bibitem [{\citenamefont {L\'evy}(1884)}]{Levy1884}%
  \BibitemOpen
  \bibfield  {author} {\bibinfo {author} {\bibfnamefont {M.~M.}\ \bibnamefont
  {L\'evy}},\ }\href@noop {} {\bibfield  {journal} {\bibinfo  {journal} {J Math
  Pure Appl},\ }\textbf {\bibinfo {volume} {10}},\ \bibinfo {pages} {5}
  (\bibinfo {year} {1884})}\BibitemShut {NoStop}%
\bibitem [{\citenamefont {Tadjbakhsh}\ and\ \citenamefont
  {Odeh}(1967)}]{Tadjbakhsh1967}%
  \BibitemOpen
  \bibfield  {author} {\bibinfo {author} {\bibfnamefont {I.}~\bibnamefont
  {Tadjbakhsh}}\ and\ \bibinfo {author} {\bibfnamefont {F.}~\bibnamefont
  {Odeh}},\ }\href@noop {} {\bibfield  {journal} {\bibinfo  {journal} {J Math
  Anal Appl},\ }\textbf {\bibinfo {volume} {18}},\ \bibinfo {pages} {59}
  (\bibinfo {year} {1967})}\BibitemShut {NoStop}%
\bibitem [{\citenamefont {Flaherty}\ \emph {et~al.}(1972)\citenamefont
  {Flaherty}, \citenamefont {Keller},\ and\ \citenamefont
  {Rubinow}}]{Flaherty1972}%
  \BibitemOpen
  \bibfield  {author} {\bibinfo {author} {\bibfnamefont {J.}~\bibnamefont
  {Flaherty}}, \bibinfo {author} {\bibfnamefont {J.}~\bibnamefont {Keller}}, \
  and\ \bibinfo {author} {\bibfnamefont {S.}~\bibnamefont {Rubinow}},\
  }\href@noop {} {\bibfield  {journal} {\bibinfo  {journal} {SIAM Journal on
  Applied Mathematics},\ }\textbf {\bibinfo {volume} {23}},\ \bibinfo {pages}
  {446} (\bibinfo {year} {1972})}\BibitemShut {NoStop}%
\bibitem [{\citenamefont {Arreaga}\ \emph {et~al.}(2002)\citenamefont
  {Arreaga}, \citenamefont {Capovilla}, \citenamefont {Chryssomalakos},\ and\
  \citenamefont {Guven}}]{Arreaga2002}%
  \BibitemOpen
  \bibfield  {author} {\bibinfo {author} {\bibfnamefont {G.}~\bibnamefont
  {Arreaga}}, \bibinfo {author} {\bibfnamefont {R.}~\bibnamefont {Capovilla}},
  \bibinfo {author} {\bibfnamefont {C.}~\bibnamefont {Chryssomalakos}}, \ and\
  \bibinfo {author} {\bibfnamefont {J.}~\bibnamefont {Guven}},\ }\href@noop {}
  {\bibfield  {journal} {\bibinfo  {journal} {Phys Rev E},\ }\textbf {\bibinfo
  {volume} {65}},\ \bibinfo {pages} {031801} (\bibinfo {year}
  {2002})}\BibitemShut {NoStop}%
\bibitem [{\citenamefont {Vassilev}\ \emph {et~al.}(2008)\citenamefont
  {Vassilev}, \citenamefont {Djondjorov},\ and\ \citenamefont
  {Mladenov}}]{Vassilev2008}%
  \BibitemOpen
  \bibfield  {author} {\bibinfo {author} {\bibfnamefont {V.}~\bibnamefont
  {Vassilev}}, \bibinfo {author} {\bibfnamefont {P.}~\bibnamefont
  {Djondjorov}}, \ and\ \bibinfo {author} {\bibfnamefont {I.}~\bibnamefont
  {Mladenov}},\ }\href@noop {} {\bibfield  {journal} {\bibinfo  {journal} {J
  Phys A-Math Gen},\ }\textbf {\bibinfo {volume} {41}},\ \bibinfo {pages}
  {435201} (\bibinfo {year} {2008})}\BibitemShut {NoStop}%
\bibitem [{\citenamefont {Djondjorov}\ \emph {et~al.}(2011)\citenamefont
  {Djondjorov}, \citenamefont {Vassilev},\ and\ \citenamefont
  {Mladenov}}]{Djondjorov2011}%
  \BibitemOpen
  \bibfield  {author} {\bibinfo {author} {\bibfnamefont {P.}~\bibnamefont
  {Djondjorov}}, \bibinfo {author} {\bibfnamefont {V.}~\bibnamefont
  {Vassilev}}, \ and\ \bibinfo {author} {\bibfnamefont {I.}~\bibnamefont
  {Mladenov}},\ }\href@noop {} {\bibfield  {journal} {\bibinfo  {journal} {Int
  J Mech Sci},\ }\textbf {\bibinfo {volume} {53}},\ \bibinfo {pages} {355}
  (\bibinfo {year} {2011})}\BibitemShut {NoStop}%
\bibitem [{\citenamefont {Mora}\ \emph {et~al.}(2012)\citenamefont {Mora},
  \citenamefont {Phou}, \citenamefont {Fromental}, \citenamefont {Audoly},\
  and\ \citenamefont {Pomeau}}]{Mora2012}%
  \BibitemOpen
  \bibfield  {author} {\bibinfo {author} {\bibfnamefont {S.}~\bibnamefont
  {Mora}}, \bibinfo {author} {\bibfnamefont {T.}~\bibnamefont {Phou}}, \bibinfo
  {author} {\bibfnamefont {J.-M.}\ \bibnamefont {Fromental}}, \bibinfo {author}
  {\bibfnamefont {B.}~\bibnamefont {Audoly}}, \ and\ \bibinfo {author}
  {\bibfnamefont {Y.}~\bibnamefont {Pomeau}},\ }\href@noop {} {\bibfield
  {journal} {\bibinfo  {journal} {Phys Rev E},\ }\textbf {\bibinfo {volume}
  {86}},\ \bibinfo {pages} {026119} (\bibinfo {year} {2012})}\BibitemShut
  {NoStop}%
\bibitem [{\citenamefont {Engler}\ \emph {et~al.}(2004)\citenamefont {Engler},
  \citenamefont {Bacakova}, \citenamefont {Newman}, \citenamefont {Hategan},
  \citenamefont {Griffin},\ and\ \citenamefont {Discher}}]{Engler2004}%
  \BibitemOpen
  \bibfield  {author} {\bibinfo {author} {\bibfnamefont {A.}~\bibnamefont
  {Engler}}, \bibinfo {author} {\bibfnamefont {L.}~\bibnamefont {Bacakova}},
  \bibinfo {author} {\bibfnamefont {C.}~\bibnamefont {Newman}}, \bibinfo
  {author} {\bibfnamefont {A.}~\bibnamefont {Hategan}}, \bibinfo {author}
  {\bibfnamefont {M.}~\bibnamefont {Griffin}}, \ and\ \bibinfo {author}
  {\bibfnamefont {D.}~\bibnamefont {Discher}},\ }\href@noop {} {\bibfield
  {journal} {\bibinfo  {journal} {Biophysical Journal},\ }\textbf {\bibinfo
  {volume} {86}},\ \bibinfo {pages} {617} (\bibinfo {year} {2004})}\BibitemShut
  {NoStop}%
\bibitem [{\citenamefont {Wakatsuki}\ \emph {et~al.}(2003)\citenamefont
  {Wakatsuki}, \citenamefont {Wysolmerski},\ and\ \citenamefont
  {Elson}}]{Wakatsuki2003}%
  \BibitemOpen
  \bibfield  {author} {\bibinfo {author} {\bibfnamefont {T.}~\bibnamefont
  {Wakatsuki}}, \bibinfo {author} {\bibfnamefont {R.}~\bibnamefont
  {Wysolmerski}}, \ and\ \bibinfo {author} {\bibfnamefont {E.}~\bibnamefont
  {Elson}},\ }\href@noop {} {\bibfield  {journal} {\bibinfo  {journal} {J Cell
  Sci},\ }\textbf {\bibinfo {volume} {116}},\ \bibinfo {pages} {1617} (\bibinfo
  {year} {2003})}\BibitemShut {NoStop}%
\bibitem [{\citenamefont {Gray}(1997)}]{Gray1997}%
  \BibitemOpen
  \bibfield  {author} {\bibinfo {author} {\bibfnamefont {A.}~\bibnamefont
  {Gray}},\ }\href@noop {} {\emph {\bibinfo {title} {Modern differential
  geometry of curves and surfaces with mathematica}}}\ (\bibinfo  {publisher}
  {CRC-Press, Boca Raton, FL},\ \bibinfo {year} {1997})\BibitemShut {NoStop}%
\bibitem [{\citenamefont {Ghibaudo}\ \emph {et~al.}(2008)\citenamefont
  {Ghibaudo}, \citenamefont {Saez}, \citenamefont {Trichet}, \citenamefont
  {Xayaphoummine}, \citenamefont {Browaeys}, \citenamefont {Silberzan},
  \citenamefont {Buguin},\ and\ \citenamefont {Ladoux}}]{Ghibaudo2008}%
  \BibitemOpen
  \bibfield  {author} {\bibinfo {author} {\bibfnamefont {M.}~\bibnamefont
  {Ghibaudo}}, \bibinfo {author} {\bibfnamefont {A.}~\bibnamefont {Saez}},
  \bibinfo {author} {\bibfnamefont {L.}~\bibnamefont {Trichet}}, \bibinfo
  {author} {\bibfnamefont {A.}~\bibnamefont {Xayaphoummine}}, \bibinfo {author}
  {\bibfnamefont {J.}~\bibnamefont {Browaeys}}, \bibinfo {author}
  {\bibfnamefont {P.}~\bibnamefont {Silberzan}}, \bibinfo {author}
  {\bibfnamefont {A.}~\bibnamefont {Buguin}}, \ and\ \bibinfo {author}
  {\bibfnamefont {B.}~\bibnamefont {Ladoux}},\ }\href@noop {} {\bibfield
  {journal} {\bibinfo  {journal} {Soft Matter},\ }\textbf {\bibinfo {volume}
  {4}},\ \bibinfo {pages} {1836} (\bibinfo {year} {2008})}\BibitemShut
  {NoStop}%
\bibitem [{\citenamefont {Mitrossilis}\ \emph {et~al.}(2009)\citenamefont
  {Mitrossilis}, \citenamefont {Fouchard}, \citenamefont {Guiroy},
  \citenamefont {Desprat}, \citenamefont {Rodriguez}, \citenamefont {Fabry},\
  and\ \citenamefont {Asnacios}}]{Mitrossilis2009}%
  \BibitemOpen
  \bibfield  {author} {\bibinfo {author} {\bibfnamefont {D.}~\bibnamefont
  {Mitrossilis}}, \bibinfo {author} {\bibfnamefont {J.}~\bibnamefont
  {Fouchard}}, \bibinfo {author} {\bibfnamefont {A.}~\bibnamefont {Guiroy}},
  \bibinfo {author} {\bibfnamefont {N.}~\bibnamefont {Desprat}}, \bibinfo
  {author} {\bibfnamefont {N.}~\bibnamefont {Rodriguez}}, \bibinfo {author}
  {\bibfnamefont {B.}~\bibnamefont {Fabry}}, \ and\ \bibinfo {author}
  {\bibfnamefont {A.}~\bibnamefont {Asnacios}},\ }\href@noop {} {\bibfield
  {journal} {\bibinfo  {journal} {Proc Natl Acad Sci USA},\ }\textbf {\bibinfo
  {volume} {106}},\ \bibinfo {pages} {18243} (\bibinfo {year}
  {2009})}\BibitemShut {NoStop}%
\bibitem [{\citenamefont {James}\ \emph {et~al.}(2008)\citenamefont {James},
  \citenamefont {Goluch}, \citenamefont {Hu}, \citenamefont {Liu},\ and\
  \citenamefont {Mrksich}}]{James2008}%
  \BibitemOpen
  \bibfield  {author} {\bibinfo {author} {\bibfnamefont {J.}~\bibnamefont
  {James}}, \bibinfo {author} {\bibfnamefont {E.~D.}\ \bibnamefont {Goluch}},
  \bibinfo {author} {\bibfnamefont {H.}~\bibnamefont {Hu}}, \bibinfo {author}
  {\bibfnamefont {C.}~\bibnamefont {Liu}}, \ and\ \bibinfo {author}
  {\bibfnamefont {M.}~\bibnamefont {Mrksich}},\ }\href@noop {} {\bibfield
  {journal} {\bibinfo  {journal} {Cell Motil Cytoskel},\ }\textbf {\bibinfo
  {volume} {65}},\ \bibinfo {pages} {841} (\bibinfo {year} {2008})}\BibitemShut
  {NoStop}%
\bibitem [{\citenamefont {Love}(1927)}]{Love1927}%
  \BibitemOpen
  \bibfield  {author} {\bibinfo {author} {\bibfnamefont {A.~E.~H.}\
  \bibnamefont {Love}},\ }\href@noop {} {\emph {\bibinfo {title} {A {T}reatise
  on the {M}athematical {T}heory of {E}lasticity}}},\ \bibinfo {edition} {4th}\
  ed.\ (\bibinfo  {publisher} {Cambridge University Press},\ \bibinfo {address}
  {Cambridge},\ \bibinfo {year} {1927})\BibitemShut {NoStop}%
\bibitem [{\citenamefont {Biot}(1965)}]{Biot1965}%
  \BibitemOpen
  \bibfield  {author} {\bibinfo {author} {\bibfnamefont {M.~A.}\ \bibnamefont
  {Biot}},\ }\href@noop {} {\emph {\bibinfo {title} {Mechanics of incremental
  deformations}}}\ (\bibinfo  {publisher} {John Wiley and Sons, New York},\
  \bibinfo {year} {1965})\BibitemShut {NoStop}%
\bibitem [{\citenamefont {Hohlfeld}\ and\ \citenamefont
  {Mahadevan}(2011)}]{Hohlfeld2011}%
  \BibitemOpen
  \bibfield  {author} {\bibinfo {author} {\bibfnamefont {E.}~\bibnamefont
  {Hohlfeld}}\ and\ \bibinfo {author} {\bibfnamefont {L.}~\bibnamefont
  {Mahadevan}},\ }\href@noop {} {\bibfield  {journal} {\bibinfo  {journal}
  {Phys Rev Lett},\ }\textbf {\bibinfo {volume} {106}},\ \bibinfo {pages}
  {105702} (\bibinfo {year} {2011})}\BibitemShut {NoStop}%
\bibitem [{\citenamefont {Hohlfeld}\ and\ \citenamefont
  {Mahadevan}(2012)}]{Hohlfeld2012}%
  \BibitemOpen
  \bibfield  {author} {\bibinfo {author} {\bibfnamefont {E.}~\bibnamefont
  {Hohlfeld}}\ and\ \bibinfo {author} {\bibfnamefont {L.}~\bibnamefont
  {Mahadevan}},\ }\href@noop {} {\bibfield  {journal} {\bibinfo  {journal}
  {Phys Rev Lett},\ }\textbf {\bibinfo {volume} {109}},\ \bibinfo {pages}
  {025701} (\bibinfo {year} {2012})}\BibitemShut {NoStop}%
\bibitem [{\citenamefont {Tallinen}\ \emph {et~al.}(2013)\citenamefont
  {Tallinen}, \citenamefont {Biggins},\ and\ \citenamefont
  {Mahadevan}}]{Tallinen2012}%
  \BibitemOpen
  \bibfield  {author} {\bibinfo {author} {\bibfnamefont {T.}~\bibnamefont
  {Tallinen}}, \bibinfo {author} {\bibfnamefont {J.~S.}\ \bibnamefont
  {Biggins}}, \ and\ \bibinfo {author} {\bibfnamefont {L.}~\bibnamefont
  {Mahadevan}},\ }\Doi {10.1103/PhysRevLett.110.024302} {\bibfield  {journal}
  {\bibinfo  {journal} {Phys Rev Lett},\ }\textbf {\bibinfo {volume} {110}},\
  \bibinfo {pages} {024302} (\bibinfo {year} {2013})}\BibitemShut {NoStop}%
\bibitem [{\citenamefont {Banerjee}\ and\ \citenamefont
  {Marchetti}(2012)}]{Banerjee2012}%
  \BibitemOpen
  \bibfield  {author} {\bibinfo {author} {\bibfnamefont {S.}~\bibnamefont
  {Banerjee}}\ and\ \bibinfo {author} {\bibfnamefont {M.~C.}\ \bibnamefont
  {Marchetti}},\ }\href@noop {} {\bibfield  {journal} {\bibinfo  {journal}
  {Phys Rev Lett},\ }\textbf {\bibinfo {volume} {109}},\ \bibinfo {pages}
  {108101} (\bibinfo {year} {2012})}\BibitemShut {NoStop}%
\bibitem [{\citenamefont {Greenhill}(1892)}]{Greenhill1892}%
  \BibitemOpen
  \bibfield  {author} {\bibinfo {author} {\bibfnamefont {A.}~\bibnamefont
  {Greenhill}},\ }\href@noop {} {\emph {\bibinfo {title} {The applications of
  elliptic functions}}}\ (\bibinfo  {publisher} {Macmillan and Co.},\ \bibinfo
  {year} {1892})\BibitemShut {NoStop}%
\bibitem [{\citenamefont {Davis}(2010)}]{Davis2010}%
  \BibitemOpen
  \bibfield  {author} {\bibinfo {author} {\bibfnamefont {H.}~\bibnamefont
  {Davis}},\ }\href@noop {} {\emph {\bibinfo {title} {Introduction to nonlinear
  differential and integral equations}}}\ (\bibinfo  {publisher} {Dover
  publications, Mineola NY},\ \bibinfo {year} {2010})\BibitemShut {NoStop}%
\end{thebibliography}%

\end{document}